\documentclass[
 aps,
 prl,
 twocolumn,
 floats,
 epsf,
 amsmath,amssymb,
 superscriptaddress,
]{revtex4-2}

\usepackage{graphicx}
\usepackage{dcolumn}
\usepackage{bm}
\usepackage{hyperref}
\usepackage[capitalise]{cleveref}
\usepackage{bbold}
\usepackage{MnSymbol}
\usepackage{physics}
\usepackage{extarrows}
\usepackage[normalem]{ulem}
\usepackage{cancel}
\usepackage{placeins}
\usepackage{booktabs}
\usepackage{soul}
\usepackage{orcidlink}

\usepackage{colortbl}
\usepackage{xcolor}
\usepackage{microtype}
\crefrangeformat{equation}{Eqs.~(#3#1#4) to (#5#2#6)}
\crefformat{equation}{Eq.~(#2#1#3)}
\crefmultiformat{equation}{Eqs.~(#2#1#3)}
{,~(#2#1#3)}{,(#2#1#3)}{,~(#2#1#3)}
\crefformat{table}{Tab.~#2#1#3}
\crefmultiformat{table}{Tabs.~#2#1#3}
{,~#2#1#3}{,#2#1#3}{,~#2#1#3}
\crefrangeformat{table}{Tabs.~#3#1#4 to #5#2#6}
\crefformat{figure}{Fig.~#2#1#3}
\crefmultiformat{figure}{Figs.~#2#1#3}
{,~#2#1#3}{,#2#1#3}{~#2#1#3}
\crefrangeformat{figure}{Figs.~#3#1#4 to #5#2#6}

\usepackage{comment}

\newcommand{\TUVienna}
{\affiliation{Institute of Solid State Physics, TU Wien, 1040 Vienna, Austria}}
\newcommand{\Kings}
{\affiliation{Physics Department, King’s College London, Strand, London WC2R 2LS, UK}}

\begin{document}

\title{Origin of misleading convergence in self-consistent many-electron theories: \\Fundamental aspects and practical implications}

\author{Herbert E{\ss}l\orcidlink{0009-0005-9883-8104}}\email{herbert.essl@tuwien.ac.at}
\TUVienna
\author{Matthias Reitner\orcidlink{0000-0002-2529-0847}}
\TUVienna
\author{Evgeny Kozik\orcidlink{0000-0001-6580-9570}}
\Kings
\author{Alessandro Toschi\orcidlink{0000-0001-5669-3377}}
\TUVienna

\date{\today}

\begin{abstract}
Self-consistent approaches in many-electron problems typically converge to an unphysical solution in strongly correlated regimes.
By deriving the mathematical condition for 
the stability of the physical solution, we unveil the \emph{precise relation} between two \emph{distinct} issues previously considered equivalent: the misleading convergence in self-consistent schemes and the multivaluedness of the Luttinger-Ward functional.
Although these problems are fundamentally linked through the divergences of the irreducible vertex function, we show that misleading convergence can occur even in the absence of such divergences. 
Eventually, a systematic procedure for stabilizing the physical solution is proposed.
\end{abstract}

\maketitle
\textsl{Introduction -- } 
Many fascinating phenomena in condensed matter physics, such as unusual spectral features (pseudogap \cite{timusk1999,vishik2018}, waterfalls \cite{graf2007,krsnik2025}, kinks \cite{byczuk2007},  etc.), unconventional superconductivity \cite{keimer2015,scalapino2012,si2023,zhou2025}, quantum criticality \cite{coleman2005,lohneysen2007,brando2016,adlerfus2024}, and Mott/Hund's metal-insulator transitions \cite{imada1998,demedici2011,isidori2019,springer2020a} are driven by electronic correlations. This sets, for their theoretical description, the high bar of solving a full quantum many-body problem of interacting fermions.
While a Feynman diagrammatic formalism 
could be employed, this lacks, unlike in QED, a small expansion parameter.  
Thus, for describing the interesting physics mentioned above, Feynman diagrammatic calculations often need to be performed in the {\sl non-perturbative} regime.

This problem affects
the important class of conserving Baym-Kadanoff approximations, constructed through self-consistent resummations of Feynman diagrams, which guarantee the thermodynamic ``consistency'' of the results \cite{baym1961,baym1962,dedominicis1964,chitra2001,potthoff2003,potthoffm.2006}.
Formally, self-consistent diagrammatic methods compute the electron Green's function $G$ as a functional $f$ of $G$ itself, with the implicit equation $G=f[G]$ being solved by iterations. With the exception of extremely simple models
the exact map $f$ is unknown, but can be constructed approximately, e.g.~as truncated diagrammatic series in terms of $G$, summed to high order by the diagrammatic Monte Carlo technique~\cite{VanHoucke2010, Kozik2010}, 
the GW scheme \cite{hedin1965,aryasetiawan1998}, or, non-perturbatively, using dynamical mean-field theory \cite{georges1996a} and its extensions \cite{maier2005,rohringer2018a}, nested-cluster \cite{vucicevic2018}  or parquet-based \cite{bickers2004a} approaches.

In the last decade, an increasing number of studies have reported that an iterative solution of $G=f[G]$ can converge to a wrong answer \cite{tandetzky2015,vucicevic2018,kozik2015,stan2015a,schafer2016,badr2024,rohshap2025}, typically in the most interesting correlated regimes. 
These studies have ascribed the occurrence of misleading convergence in the iterative solution for $G$ to the intrinsic multivaluedness of the Luttinger-Ward functional (LWF) of the many-electron problem \cite{rossi2015,kim2020a,kozik2015,potthoffm.2006}.  
Any crossing of two branches of the LWF (with only one being physical)~\footnote{Precisely, it was proven that the crossing of two solutions of the equation $G[G_0]=G_\text{phys}$ implies a vertex divergence. 
However, the existence of multiple solutions for this equation originates from the multivaluedness of the LWF.} implies the simultaneous divergence of the two-particle irreducible (2PI) vertex function \cite{schafer2013,janis2014,schafer2016,gunnarsson2016a,vucicevic2018,chalupa2018,thunstrom2018a,springer2020,pelz2023a,essl2024}. Physically, this divergence is intrinsically linked \cite{gunnarsson2016a,chalupa2021a,mazitov2022,adler2024} to one of the most relevant feats of electronic correlation: the local magnetic moment formation and the associated freezing of charge fluctuations, as this huge differentiation of on-site magnetic and charge responses requires extremely large vertex functions. Notably, vertex divergences also yield the necessary condition \cite{reitner2020,reitner2024,kowalski2024,meixner2025,moghadas2026} for the onset of phase-separations in the proximity of Mott-Hubbard metal-insulator transitions \cite{imada1998,georges1996,kotliar2002,eckstein2007}. 

More formally, vertex divergences represent a key hallmark \cite{gunnarsson2017,reitner2020,chalupa2021a,melnick2020,mazitov2022,mazitov2022a,adler2024} of the breakdown of self-consistent perturbation theory (SCPT). 
Specifically, the skeleton diagrammatic series in terms of $G$ was found~\cite{kozik2015, vanhoucke2024, kim2022a} to always converge to an unphysical answer for interaction values larger than those at which the first branching point of the LWF and, hence, the first vertex divergence, occurs.
However, so far, the evidence for the connection between the misleading convergence of the iterative solution of $G=f[G]$ and the crossings between the LWF branches has remained \emph{empirical}. 
This is likely the reason, why the misleading convergence issues  affecting GW \cite{stan2015a}, parquet-based \cite{badr2024}, and nested-cluster schemes \cite{vucicevic2018}, often in the proximity of vertex divergences,
have been hitherto attributed to the branching of the LWF, even if these iterative approaches, differently from SCPT, are \emph{not} based on perturbative series of LWF diagrams.
At the same time, misleading convergences have also been reported in regions of parameter space \emph{without} such crossings, i.e.~vertex divergencies \cite{kozik2015,vucicevic2018,kim2022a}. Evidently, a complete understanding of this problem and its implications is still lacking. 

\begin{figure}[t!]
\hspace{-10mm}
\includegraphics[scale=0.4]{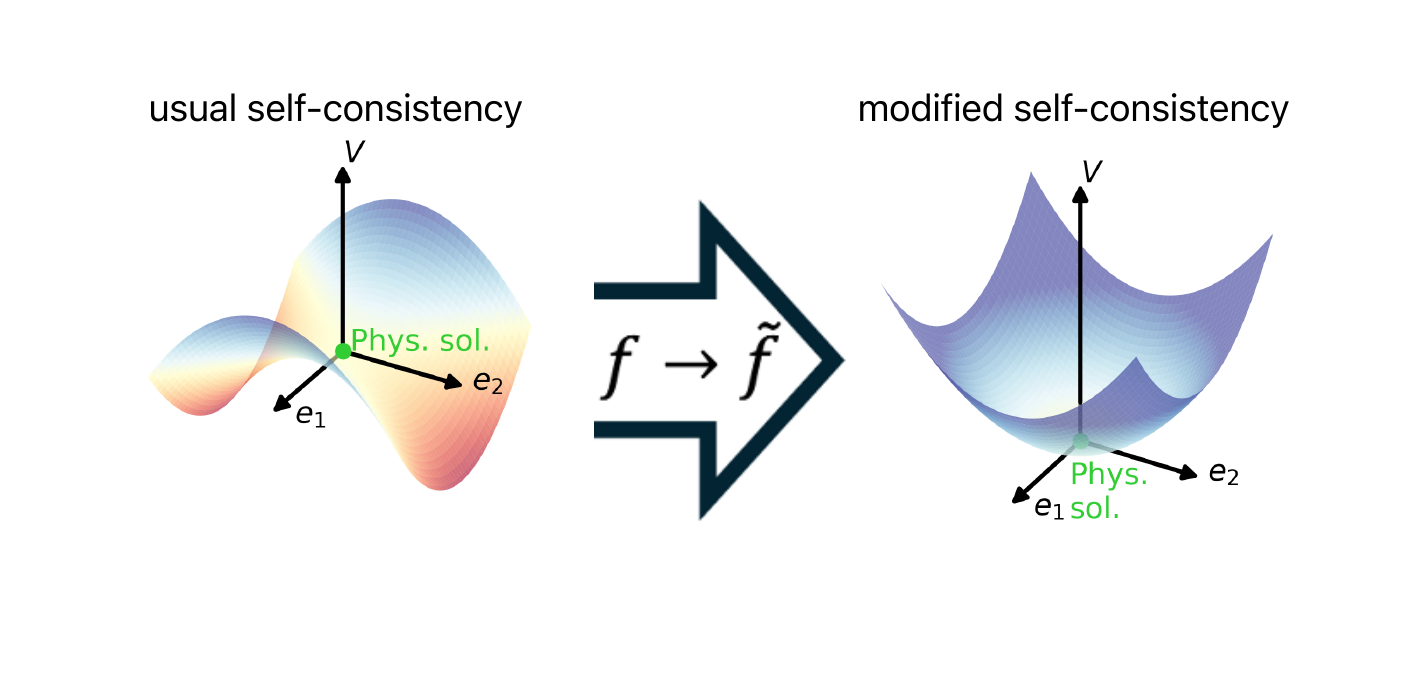}
\vspace{-15mm}
\caption{\label{fig:schematic} Heuristic sketch of the "stability landscape" of a self-consistent scheme. 
Left: Usual self-consistent scheme ($f$) in a parameter regime, where the physical fixed point is unstable in one direction ($e_2$).
Right: Modified self-consistent scheme ($\tilde f$), where the physical fixed point has been stabilized.} 
\end{figure}

In this work, we aim to clarify the origin of the misleading convergence observed in iterative many-body schemes, formally \emph{disentangling} this issue from the branching of the LWF. To this end, we first devise a rigorous, model-independent procedure to determine, whether the fixed points of the map $f$ are stable or unstable. 
This allows us to prove that,
for the same parameters at which branch crossings of the LWF (and the associated vertex divergences) occur, the physical fixed point of the iterative procedure also becomes unstable. However, as we will also demonstrate, in absence of vertex divergences, both the misleading convergence and the breakdown of the skeleton series can still occur but \emph{no longer simultaneously}.
Through our analytical derivation, we will be able to eventually unveil the \emph{precise relation} between these two important issues of the self-consistent theories for the many-electron problem.
Finally, we derive a systematic approach for changing the map $f$ (to a \emph{modified} map $\tilde{f}$) that stabilizes the physical fixed point (see the illustration in \cref{fig:schematic}), thereby ensuring convergence to the correct physical solution over the entire parameter range.

\textsl{Models -- } Our analytic theory will be illustrated by its application to the zero point (ZP) model, for which the functional $f$ simplifies to a function, and to the Hubbard atom (HA). These are defined by the action
\begin{align}
\label{eq:S_general}
    S = -\sum_{\nu,\sigma} \bar c_{\nu,\sigma} G_{0,\nu}^{-1}  c_{\nu,\sigma} + S_\text{int},
\end{align}
where $\nu$ is a fermionic Matsubara frequency, $\bar c(c)$ are Grassmann variables that represent the creation (annihilation) of an electron (see appendix A) and $G_{0,\nu}^{-1}=i\nu+\delta\mu$ with $\delta\mu$ being the chemical potential ($\delta\mu=0$ corresponds to ph-symmetry). 
The interacting part of the action reads $S_\text{int,ZP}=\frac{U}{\beta}\sum_\nu \bar c_{\nu,\uparrow}c_{\nu,\uparrow} \bar c_{\nu,\downarrow}c_{\nu,\downarrow}$ for the ZP model and $S_\text{int,HA}=U \int_0^\beta d\tau\, (\bar c_{\tau,\uparrow}c_{\tau,\uparrow}-1/2) (\bar c_{\tau,\downarrow}c_{\tau,\downarrow}-1/2)$ for the HA, with $U$ being the on-site interaction, $\beta$ the inverse temperature and $\tau$ the imaginary time.
 
These basic models are selected for illustrative purposes and for a rigorous benchmark w.r.t.~their exact solutions. Yet, they still entail, in a crude fashion, pivotal feats of the electron-correlation physics. The HA, which describes totally localized electrons, yields a somewhat ``extreme representation'' of Mott insulating regimes. As for the ZP, at half-filling, it essentially corresponds to a binary-mixture disorder model, precisely its coherent potential approximation (CPA) solution in the limit of high-lattice coordination \cite{schafer2016} (see SM~\cite{SM}). This features a \emph{metal-to-insulator} transition driven by increasing disorder-strength (which corresponds to $U$ in the ZP), often used pedagogically to provide an intuition  for the more complex Mott transition. Thus, not so surprisingly, both HA and ZP are plagued by qualitatively similar misleading convergence issues \cite{kozik2015,stan2015a,rossi2015,janis2014,schafer2016,gunnarsson2017,essl2024} 
as more realistic Hubbard-like models \cite{schafer2013,schafer2016,gunnarsson2016a,gunnarsson2017,chalupa2021a,vucicevic2018,pelz2023a,meixner2025}.

\textsl{Self-consistent schemes --} In previous literature two different iterative schemes have been exploited to investigate possible misleading convergence issues. 
Both use a damped iteration scheme together with the Dyson equation, which links the non-interacting Green's function ($G_0$) to the interacting one ($G$) via the self-energy $\Sigma$, i.e.,~$G^{-1}=G_0^{-1}-\Sigma$. The two schemes differ, however, for the quantity that is calculated by the iterative procedure, i.e. either ~(I)
$G_0$ \cite{kozik2015} or~(II) $G$ \cite{baym1962,abrikosov1975a,martin2016}.
While the two approaches have provided essentially equivalent indicators for studying misleading convergence problems, for the purposes of our investigation, it is important to precisely work out their intrinsic differences. Specifically, in scheme (I) $G$ is fixed to the physical Green's function $G_\text{phys}$ and $G_0$ is found iteratively using the (unique) self-energy functional $\Sigma[G_0]$ 
\begin{align}
\label{eq:it_g0}
\begin{split}
    &\left[G_{0,\nu}^{(n+1)}\right]^{-1} = f_{G_0,\nu}\left[\left[G_{0}^{(n)}\right] ^{-1}\right] \; \;\text{with:} \\
    & f_{G_0,\nu}\left[G_{0}^{-1}\right] = p\,\left(G_{\text{phys},\nu}^{-1}+\Sigma_\nu[G_0] \right) +(1-p)\, G_{0,\nu}^{-1},\\
\end{split}
\end{align}
where $n$ is the iteration index
and the iteration damping $p\in(0,1]$. For simplicity, we omit the momentum, spin and orbital dependence, but the generalization is straightforward (see Appendix B). In scheme (II),
which corresponds to the conventional SCPT at weak coupling, $G_0$ is fixed to the physical non-interacting Green's function $G_{0,\text{phys}}$ and $G$ is iterated using the self-energy functional $\Sigma[G]$ of the LWF-formalism:
\begin{align}
\label{eq:it_g}
\begin{split}
    &\left[G_{\nu}^{(n+1)}\right]^{-1} = f_{G,\nu}\left[\left[G^{(n)}\right] ^{-1}\right] \; \;\text{with:} \\
    &f_{G,\nu}\left[G^{-1}\right] = p\,\left(G_{0,\text{phys},\nu}^{-1}-\Sigma_{\nu}[G] \right) +(1-p)G_{\nu}^{-1}.
\end{split}
\end{align}
The solution of \cref{eq:it_g0,eq:it_g} corresponds to a stable fixed point of the respective iterative map, $f_{G_0}$ or $f_G$.\\
Both schemes are affected by the intrinsic multivaluedness of the LWF \cite{kozik2015,vucicevic2018,kim2020a}, albeit in different ways. 
In scheme (I) one finds {\sl several} $G_0$ functions that fulfill the condition $G[G_0]=(G_0^{-1}-\Sigma[G_0])^{-1}=G_\text{phys}$ \cite{kozik2015,gunnarsson2017}. 
Therefore, \cref{eq:it_g0} has multiple fixed points, of which only one corresponds to $G_{0, \text{phys}}$.

For scheme (II), instead, the multivaluedness is directly reflected onto the non-uniqueness of the functional $\Sigma[G]$, i.e., unlike $\Sigma[G_0]$, several functionals $\Sigma[G]$ exist, while only one of them (depending on the parameter regime) describes the physical self-energy \cite{kozik2015, rossi2015,schafer2016,vanhoucke2024}. Hence, $G_\text{phys}$ is only a fixed point of \cref{eq:it_g} if the correct branch of $\Sigma[G]$ is chosen.
\begin{figure*}[th!]
\includegraphics[width=\textwidth]{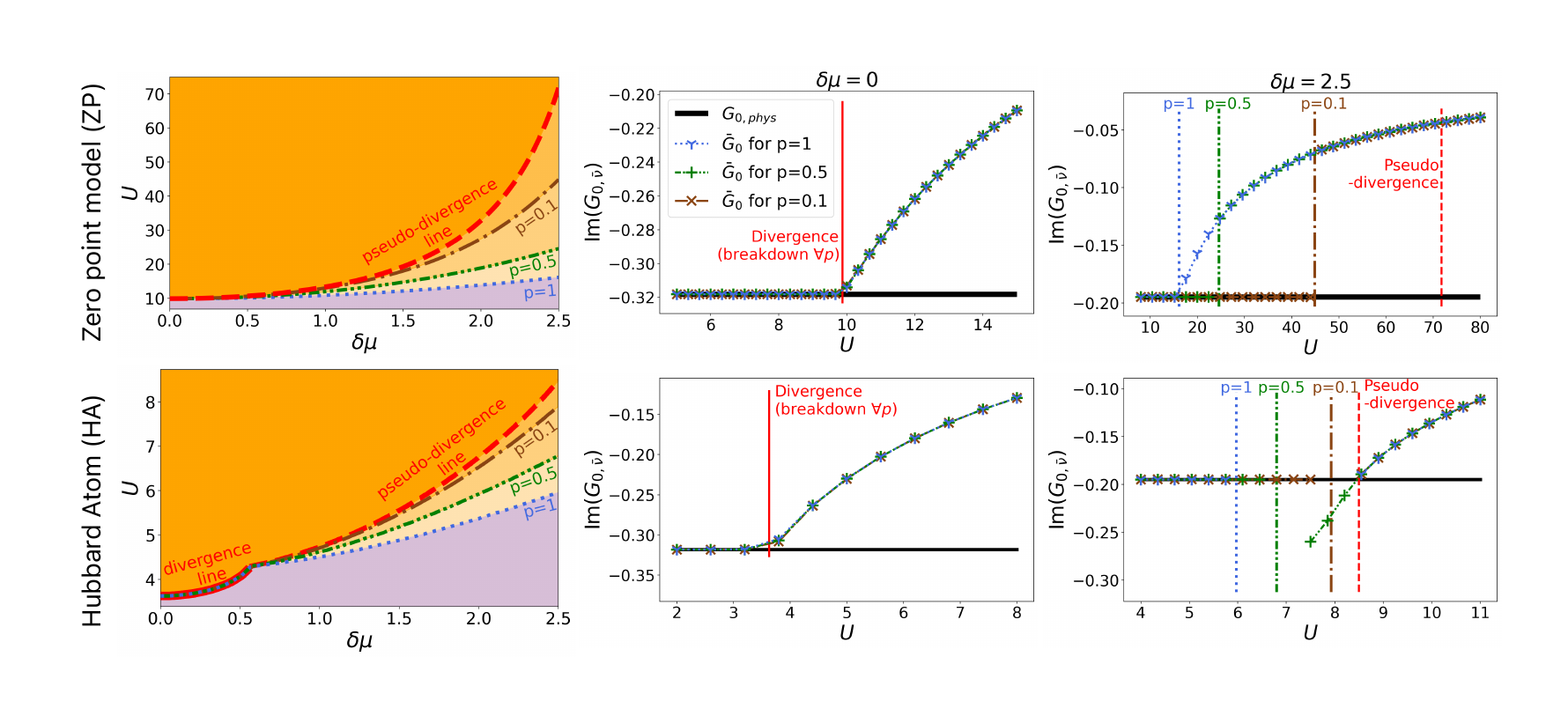}
\vspace{-15mm}
\caption{\label{fig:misleading_g0} Misleading convergence for the scheme of \cref{eq:it_g0} in the ZP model (upper panels) and HA (lower panels). Left panels: Phase space of both models, where the lines corresponding to the instability condition of the iterative scheme are shown for different dampings together with the (pseudo-)divergence lines (red solid/dashed lines). 
The imaginary part of $\bar G_{0,\bar\nu=\pi/\beta}$ obtained via iteration of \cref{eq:it_g0} is 
compared to the physical solution $G_{0,\text{phys},\bar\nu}$ at/out ph-symmetry (middle/right panels).}
\end{figure*}

\textsl{Iteration in $G_0$ --} We start by explicitly considering the map $f_{G_0}$ in \cref{eq:it_g0}.
In the spirit of Ref.~\cite{vanloon2020a}, we note that the information about the stability of the fixed points of $f_{G_0}$ is encoded in its Jacobian $J_{G_0}$~\footnote{Rigorously, the Jacobian is sufficient to determine the stability of a fixed point if the latter is hyperbolic, i.e.,~the Jacobian has no eigenvalue with unit modulus~\cite{guckenheimer1983}.}.
Specifically, a fixed point is stable 
if the Jacobian evaluated on the given fixed point has only eigenvalues with magnitudes smaller than 1 (see e.g. \cite{strogatz2015,argyris2017,chen2023}).
In our case, the Jacobian $J_{G_0}$, which is a functional of $G_0^{-1}$, can be directly expressed through the generalized (static, i.e.,~ zero transfer frequency) charge susceptibility $\chi_\text{c}^{\nu\nu^\prime}$ of the system considered and its bubble term counterpart, $\chi_{0,\nu\nu^\prime}=-\beta G_\nu^2 \delta_{\nu\nu^\prime}$ (see Appendix A and C), as
\begin{align}
\label{eq:J_g0}
    J^{\nu\nu^\prime}_{G_0} = \fdv{f_{G_0,\nu}}{G_{0,\nu^\prime}^{-1}} = \delta_{\nu \nu^\prime} - p \, [\chi_0^{-1}]_{\nu\nu_1}\chi_\text{c}^{\nu_1\nu^\prime},
\end{align}
where summations are performed on repeated indices. 
The eigenvalues of the Jacobian \cref{eq:J_g0} are, thus, $1-p\lambda_\alpha$, $\lambda_\alpha$ being the eigenvalues of the matrix $\chi_0^{-1}\chi_\text{c}$.

Let us first consider the case, where the eigenvalue, which is responsible for the instability, i.e.~$|1-p\lambda_\alpha |>1$, is real. In this situation, a fixed point is unstable if an eigenvalue $\lambda_\alpha$ of $\chi_0^{-1}\chi_\text{c}$ satisfies one of two conditions: (i) $\lambda_\alpha <0$ or (ii) $\lambda_\alpha >2/p$. 
These two situations are fundamentally different: the onset of condition (i), which does {\sl not} depend on $p$, corresponds to a sign-change of $\lambda_\alpha$, and, hence, to a degenerate matrix $\chi_\text{c}$, triggering a \emph{divergence} \cite{schafer2013} of the corresponding 2PI vertex: $\Gamma_\text{c} \sim \chi_\text{c}^{-1}\chi_0$. Hence, (i) represents an intrinsic non-perturbative property of the considered system \cite{gunnarsson2016a, gunnarsson2017, reitner2020,chalupa2021a,adler2024,mazitov2022}.
On the contrary, the occurrence of the instability condition (ii)
can be systematically overcome by damping the iteration scheme, i.e., choosing a smaller $p$. Hence, (ii) should be regarded as a mere algorithmic aspect of the self-consistency. 
For particle-hole(ph)-symmetric systems, where all eigenvalues of $\chi_0^{-1}\chi_\text{c}$ are real \cite{rohringer2012a,springer2020,vanloon2020a,essl2024}, only these two types of instabilities can occur.  
However, out of ph-symmetry, where the eigenvalues of $\chi_0^{-1}\chi_\text{c}$ can become complex conjugate pairs, another type of instability appears. This instability, which we classify as of type (iii), is of particular importance: It is triggered by two complex conjugated eigenvalues of the Jacobian generically expected in the absence of ph-symmetry \cite{reitner2020,essl2024}) and appears near the change of sign of the real part of $\lambda_\alpha$.
In particular, for a complex $\lambda_\alpha$, the real part contributes linearly in $p$ to $\abs{1-p\lambda_\alpha}^2$
of $J_{G_0}$, while its imaginary part contributes only quadratically.
Hence, for $p\rightarrow0^+$, it is the sign switch of Re($\lambda_\alpha$) from positive to negative which triggers the instability. For a loose analogy with (i), the condition Re($\lambda_\alpha$)=0 with Im($\lambda_\alpha$) $\neq 0$ has been referred to in the literature \cite{essl2024,reitner2024,vucicevic2018} as \emph{pseudo-divergence} \footnote{Albeit or a slightly different quantity, i.e. for $\chi_\text{c}$, instead of $\chi_0^{-1}\chi_\text{c}$. Note that while the pseudo-divergences have algorithmic consequences, their physical interpretation remains still unclear.}. 

To illustrate the applicability  of our scheme we consider the ZP model and the HA. 
The obtained results are reported in \cref{fig:misleading_g0}, where the top row presents the data for the ZP and the bottom row for the HA model.
The left panels show the location of vertex divergences and pseudo-divergences in the $U$-$\delta\mu$-space (we chose $\beta=1$ for our units). Further, the lines beyond which the physical solution becomes unstable for a given $p$ [$J_{G_0}$ has eigenvalue(s) of magnitude 1] are plotted. Note that the ZP displays vertex divergences only at ph-symmetry, while in the HA they can appear also in a finite $\delta \mu$ range.  Above the vertex divergence and pseudo-divergence lines the physical solution is unstable at any $p$:
According to \cref{eq:J_g0}, by increasing $U$, we will get a $p$-independent [type (i)] instability of the physical fixed point exactly at the vertex divergence line. For $|\delta \mu| >0$ in the ZP model and for $|\delta\mu| > 0.6$ in the HA model, an instability of type (iii) occurs, whose location progressively approaches the pseudo-divergence line for $p\rightarrow 0^+$.
Note that instabilities of type (ii), i.e., $\lambda_\alpha >2/p$, do not appear in the studied parameter regime. They would, however, appear in the case of an attractive interaction ($U<0$). 
The numerical iterative solution \footnote{Note that while $\Sigma[G_0]$ is analytically known for the ZP model, it is not for the HA, therefore the CT-INT solver [\onlinecite{gull2011}] of \emph{TRIQS} [\onlinecite{parcollet2015}] is used to calculate $\Sigma[G_0]$ for the HA. Some of the results have also been checked by using the CT-HYB solver [\onlinecite{gull2011}] of \emph{w2dynamics} [\onlinecite{wallerberger2019}] together with the \emph{sparse-ir} library [\onlinecite{wallerberger2023}]. The physical Green's function $G_\text{phys}$ is analytically known for both models.} of \cref{eq:it_g0}, shown in the middle/right panels of \cref{fig:misleading_g0} fully confirm these predictions. 
Specifically, in the middle panels the imaginary part of the physical $G_{0,\text{phys}}$ at the first Matsubara frequency $\bar\nu = \frac{\pi}{\beta}$ is shown together with the self-consistent solution $\Bar G_{0}$ of \cref{eq:it_g0} at ph-symmetry for different $U$ values: As expected, $\Bar G_{0} \! = \!$ $G_{0,\text{phys}}$ (see SM~\cite{SM} for the real parts)  
for small $U$ up to the vertex divergence line (independent of $p$).  Afterwards, for larger $U$ values, \cref{eq:it_g0} can still be converged,  
but to an {\sl unphysical} solution ($\Bar G_{0} \! \neq \! G_{0,\text{phys}}$) \cite{kozik2015}.
In the right panels of \cref{fig:misleading_g0} the same calculations have been repeated out of ph-symmetry at $\delta\mu=2.5$, where a pseudo-divergence of $\chi_0^{-1}\chi_\text{c}$ instead of a divergence occurs. 
Consistent with our predictions, the convergence of $\Bar G_{0}$ to $G_{0,\text{phys}}$ now becomes $p$-dependent, with its hard boundary determined by the pseudo-divergence line~\footnote{Another difference is that the misleading convergence to an unphysical solution now happens discontinuously (see SM~\cite{SM}).}.

\textsl{How to stabilize the physical solution --} 
\begin{figure}[t!]
\vspace{-10mm}
\includegraphics[scale=0.8]{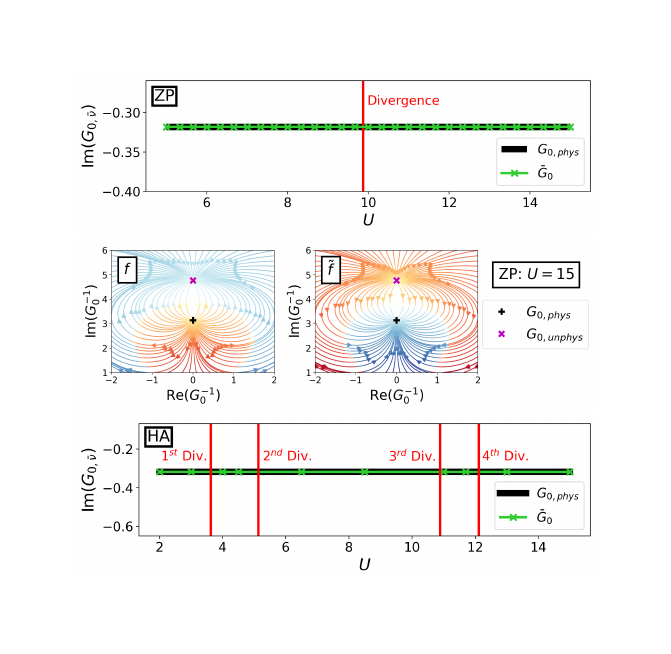}
\vspace{-15mm}
\caption{\label{fig:stay_g0}Im($G_{0,\bar\nu}$) obtained via the modified iterative scheme $\tilde f$ \cref{eq:ftilde_g0} at ph-symmetry ($\delta\mu=0, p=0.1$) for both the ZP model (top) and the HA (bottom), compared with the corresponding physical solution. Middle panel: flow-diagram in the complex plane of $G_0^{-1}$ for the ZP model in the non-perturbative regime ($U =15$) for the conventional  \cref{eq:it_g0} (left) and the modified scheme \cref{eq:ftilde_g0} (right), where stability region of a fixed point is indicated by the blue color.}
\end{figure}
To stabilize the physical solution in the entire range of parameters, we introduce the modified iterative map $\tilde f$:
\begin{align}
\label{eq:ftilde_g0}
\begin{split}
    &\left[G_{0,\nu}^{(n+1)}\right]^{-1} = \tilde f_{G_0,\nu}\left[\left[G_{0}^{(n)}\right] ^{-1}\right] \\
    &\text{with } \tilde f_{G_0,\nu}\left[G_{0}^{-1}\right]= \mathcal{P}^{\nu\nu_1}(G_{\text{phys},\nu_1}^{-1}+\Sigma_{\nu_1}[G_0] ) \\
    &\hspace{35mm}+(\delta_{\nu\nu_1}-\mathcal{P}^{\nu\nu_1}) G_{0,\nu_1}^{-1},
\end{split}
\end{align}
where $\mathcal{P}$ is defined as $\mathcal{P}^{\nu\nu^\prime}={\cal U}^{\nu\alpha_1}\mathcal{D}^{\alpha_1\alpha_2}[{\cal U}^{-1}]^{\alpha_2\nu^\prime}$, with ${\cal U}$ being the similarity transformation that diagonalizes $J_{G_0}$, \cref{eq:J_g0}, (and therefore also $\chi_0^{-1}\chi_\text{c}$) and
\begin{align} 
\label{eq:daa}
\mathcal{D}^{\alpha\alpha^\prime} = 
    \begin{cases}
      p\delta_{\alpha\alpha^\prime}, & \text{if}\ \text{Re}(\lambda_\alpha)>0 \\
      -p\delta_{\alpha\alpha^\prime}, & \text{if}\ \text{Re}(\lambda_\alpha)\leq 0
    \end{cases}  
\end{align}
Hence, $\mathcal{P}$ can be readily computed through {\sl eigenvalues} and {\sl eigenvectors} of $\chi_0^{-1}\chi_\text{c}$~\footnote{The explicit expression of $\mathcal{P}$ through the eigenvalues/eigenvectors of $\chi_0^{-1}\chi_\text{c}$ eventually provides a rigorous justification of the ad-hoc  ``scheme B'', introduced in Ref.~\cite{kozik2015} to stabilize the physical solution of the HA after the first vertex divergence.}.
The Jacobian of the modified iterative scheme, \cref{eq:ftilde_g0}, then reads
\begin{align}
    \tilde J_{G_0} = \fdv{\tilde f_{G_0,\nu}}{G_{0,\nu^\prime}^{-1}} = \delta_{\nu\nu^\prime} -\mathcal{P}^{\nu\nu_1}[\chi_0^{-1}]_{\nu_1\nu_2}\chi_\text{c}^{\nu_2\nu^\prime}. 
\end{align}
Its eigenvalues are $1-\mathcal{D}^{\alpha\alpha}\lambda_\alpha$, hence (for a small enough $p$) the Jacobian of $\tilde{f}$ has no eigenvalue of magnitude greater than 1, even after the (pseudo-)divergence lines.
Note that this procedure can be directly applied, {\sl mutatis mutandis,} to other (exact or approximated) self-consistent schemes exhibiting unstable physical fixed points.

In \cref{fig:stay_g0} the modified iterative scheme of \cref{eq:ftilde_g0} is applied to the ZP model (top) and the HA (bottom) at ph-symmetry (whereas it can be successfully exploited also out of ph-symmetry, see SM~\cite{SM}).
Note that, in practice, the knowledge of the physical fixed point is \emph{not} needed for the application of our scheme (see Appendix D and SM~\cite{SM}) and, further, that the Jacobian is \emph{not} computed during the self-consistency, but only for the last iteration.
Our results make it evident that the modified iterative scheme $\tilde f$ does actually converge to the physical solution also after the crossing of (multiple in the case of the HA \cite{schafer2016,thunstrom2018a,essl2024}) 
vertex divergence lines. 
This result is highlighted by the flow-diagrams (see Appendix E) in the middle panel of \cref{fig:stay_g0} showing the evolution of the iteration procedure towards the solution for the ZP model at $U=15$, i.e., after the vertex divergence line, both for the usual $f$ (left) and the modified  $\tilde f$ scheme (right).  
\begin{figure}[t!]
\includegraphics[scale=0.35]{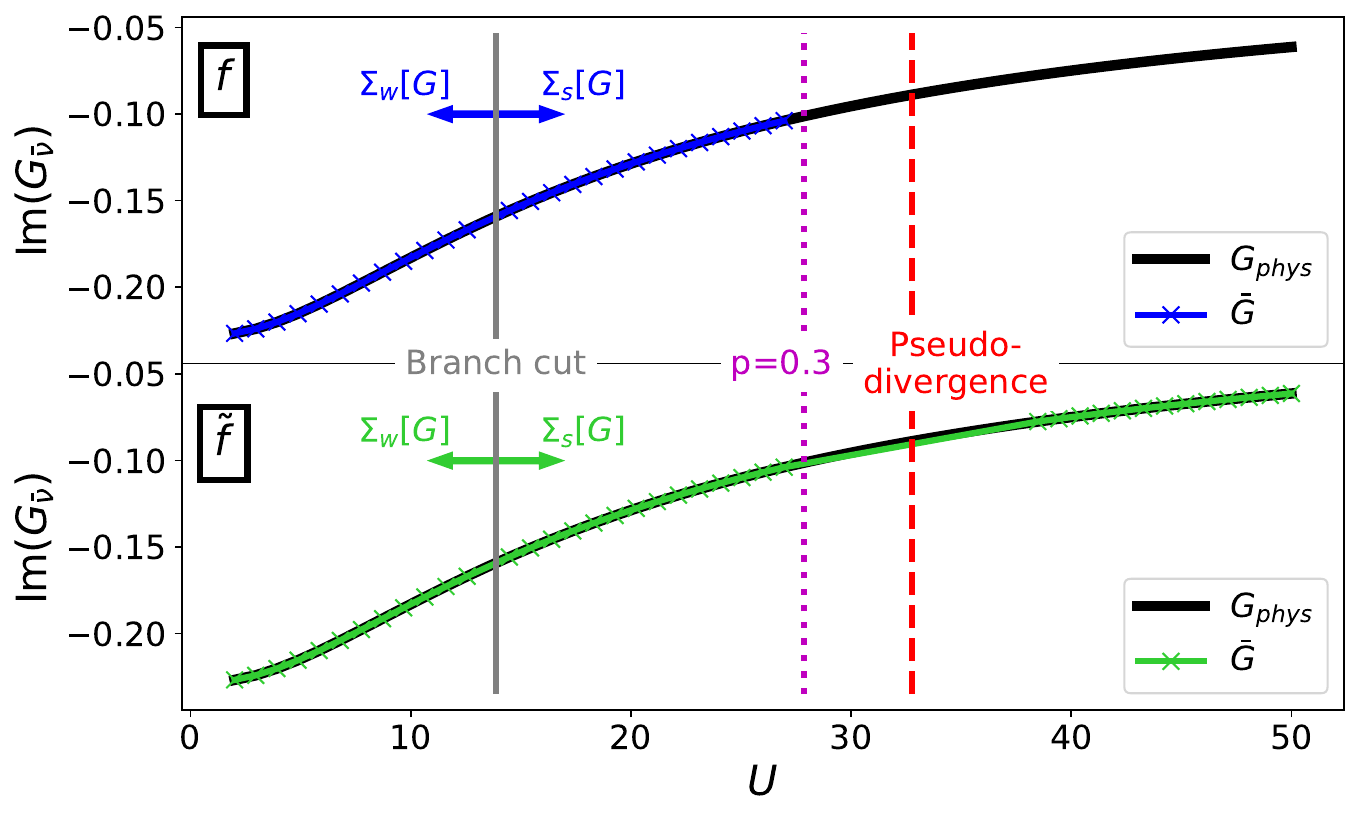}
\vspace{-5mm}
\caption{\label{fig:bold}Self-consistently calculated $G$ for the ZP model out of ph-symmetry ($\delta\mu=2, p=0.3$) with the conventional \cref{eq:it_g} (top) and the modified scheme \cref{eq:ftilde_g} (bottom), see text.} 
\vspace{-5mm}
\end{figure}

\textsl{Iteration in $G$ --} 
We turn to stabilizing the physical solution in its practical calculations via \cref{eq:it_g}, i.e., when $G_\text{phys}$ is the unknown. 
Assuming that the correct branch for $\Sigma[G]$ is chosen (i.e., $G_\text{phys}$ is indeed a fixed point of \cref{eq:it_g}), we can now calculate the Jacobian of \cref{eq:it_g}, which reads
\begin{align}
\label{eq:J_g}   J^{\nu\nu^\prime}_{G} = \fdv{f_{G,\nu}}{ G_{\nu^\prime}^{-1}} = \delta_{\nu,\nu^\prime} - p \, [\chi_\text{c}^{-1}]^{\nu\nu_1}\chi_{0,\nu_1\nu^\prime}.
\end{align}
As the eigenvalues of  $J_G$ are $1-p/\lambda_\alpha$, it is clear that the instability of the physical solution (for $p\rightarrow 0^+$) will occur at vertex divergences and pseudo-divergences of $\chi_0\chi_\text{c}^{-1}$ 
As this quantity is the inverse of the term $\chi_0^{-1}\chi_\text{c}$, relevant for the  scheme of \cref{eq:it_g0}, all pseudo-divergence lines 
have the same location in both schemes, while the vertex divergences lines now correspond to odd poles instead of zero crossing for the eigenvalues of $\chi_0\chi_\text{c}^{-1}$.  \\
In order to stabilize the physical fixed point after the corresponding (pseudo-)divergence lines, we can adopt a similar approach as before, introducing a modified iterative scheme $\tilde f_G$ \cref{eq:ftilde_g}, which is defined as
\begin{align}
\label{eq:ftilde_g}
\begin{split}
  &G^{(n+1)}_\nu = \tilde f_{G,\nu}[G^{(n)}]\\
    &\text{with } \tilde f_{G,\nu}[G]=  \mathcal{P}^{\nu\nu_1}\left(G_{\text{0,phys},\nu_1}^{-1}-\Sigma_{\nu_1}[G]\right) \\
    &\hspace{35mm}+(\delta_{\nu\nu_1}-\mathcal{P}^{\nu\nu_1}) G_{\nu_1}^{-1},
\end{split}
\end{align}
with $\mathcal{P}$ as in \cref{eq:ftilde_g0,eq:daa}.
The corresponding Jacobian is
\begin{align}
\label{eq:J_tilde_g}   
\tilde J^{\nu\nu^\prime}_{G} = \fdv{\tilde f_{G,\nu}}{ G_{\nu^\prime}^{-1}} = \delta_{\nu,\nu^\prime} - \mathcal{P}^{\nu\nu_1} \, [\chi_\text{c}^{-1}]^{\nu_1\nu_2}\chi_{0,\nu_2\nu^\prime},
\end{align}
whose eigenvalues are $1-\mathcal{D}^{\alpha\alpha}/\lambda_\alpha$. Hence, as in \cref{eq:ftilde_g0}, the Jacobian of the modified scheme will not have eigenvalues of magnitude larger than $1$ for $p \rightarrow 0^+$.\\ 
The numerical results of the calculation of $G$ in the ZP model are shown in \cref{fig:bold}, comparing the conventional iteration \cref{eq:it_g} (top) and the modified scheme \cref{eq:ftilde_g} (bottom). 
Note that in both cases, the correct branch of the $\Sigma[G]$ functional has been taken (i.e., the weak-coupling branch  $\Sigma_\text{w}$ for $U<U_\text{branch}$ (gray line in \cref{fig:bold}) and the strong-coupling one $\Sigma_\text{s}$ for $U>U_\text{branch}$, which are exactly known for the ZP model \cite{rossi2015,vanhoucke2024}). However, even if the correct functional branch has been used, the conventional scheme $f_G$ converges to physical solution only up to the ($p$-dependent) predicted instability border (cf.~\cref{fig:bold}, top panel), approaching the pseudo-divergence line for $p \rightarrow 0^+$.   
The data of the bottom panel demonstrate, instead, how the problem gets solved by using the modified scheme $\tilde{f}$: We find a convergence to the physical solution {\sl both} before and after the pseudo-divergence line\footnote{Note that we only show data for the ZP model here, as the map $f_G$ is only known perturbatively for the HA.}.

\textsl{Conclusion --}  
By means of our analysis, we were able to separate two issues, which affect self-consistent approaches for the many-electron problem, previously considered somewhat intertwined \cite{stan2015a,kozik2015,tandetzky2015,rossi2015,schafer2016, gunnarsson2017,tarantino2017,vucicevic2018,vanhoucke2024}: the convergence of the iterative procedure and the multiple branches of the LWF. 
Specifically, we introduced a \emph{model-independent} method to rigorously determine the stability of the physical solution in self-consistent approaches based on the electron Green's function, and illustrate 
the modification needed to avoid the  associated misleading convergence problems.

The separate problem of multivaluedness of the LWF reveals itself in the fact that the skeleton series for $\Sigma[G]$ always converges to the weak coupling branch of the LWF \cite{kozik2015}, despite this branch being incorrect in the strong coupling regime \cite{kehrein1998,pelz2023a}.
The origin of such unphysical convergence is \emph{not} the instability of the physical fixed point, since this also happens if $\Sigma[G]$ is evaluated at the exact Green's function $G_\text{phys}$ \cite{kozik2015}.
Despite recent progress in rectifying the SCPT instabilities caused by divergences in the series for $\Sigma[G]$~\cite{kim2022a}, capturing the strong-coupling branch(es) of the LWF through diagrammatic expansion remains an unsolved problem.
On the other hand, as the instability of the physical fixed point revealed here can occur \emph{even if} the correct branch of the LWF is used for 
$\Sigma[G]$ (\cref{fig:bold}, top panel), understanding how and at which parameters the iterative procedure should be adjusted will allow future studies to focus exclusively on the remaining problem of constructing the correct functional $\Sigma[G]$ in the non-perturbative regime.
Meanwhile, the misleading convergence issues reported for approaches such as GW, parquet, or nested-cluster-based schemes~\cite{stan2015a,badr2024,vucicevic2018}---which do not rely on an explicit resummation of the skeleton expansion (and are thus free from the problem of selecting the correct branch of the LWF)---could be directly amendable using the prescription presented here.

\begin{acknowledgments}
\textit{Acknowledgements.} 
We thank Samuel Badr, Sergio Ciuchi, Lorenzo Crippa, Marie Eder, Marcel Gievers, Emanuel Gull, Nikolaus Haunschmied, Daniel Hoffmann, Anna Kauch, Michael Meixner, Mário Malcolms de Oliveira, Stefan Rohshap, Georg Rohringer, and Giorgio Sangiovanni for insightful discussions. H. E., M. R., and A. T. acknowledge support from the Austrian Science Fund (FWF) [Grant DOI: 10.55776/ I5487]. M. R. further acknowledges support as a recipient of a DOC fellowship of the Austrian Academy of Sciences. E. K. was supported by EPSRC through Grant No. EP/ X01245X/1.
\end{acknowledgments}

\textit{Data availability --} A data set containing all numerical data and scripts for calculating and plotting these data is publicly available on the TU Wien Research Data repository \cite{data}.

\bibliography{refs}
\clearpage
\newpage

\onecolumngrid
\noindent
\centerline{\textbf{\large End Matter}}
\twocolumngrid

\textsl{Appendix A: Definition of one- and two-particle quantities --}
The one-particle Green's function in the local, SU(2)-symmetric, single-orbital case is defined as
\begin{align}
\label{eq:defG}
    G_\nu = \frac{1}{Z}\int \mathcal{D}[\Bar{c},c]\; \frac{1}{2} \sum_\sigma \bar c_{\nu,\sigma} c_{\nu,\sigma}\; e^{-S}
\end{align}
where $Z=\int \mathcal{D}[\Bar{c},c]\;e^{-S}$ and $S$ is the action that depends on the considered model [see \cref{eq:S_general}]. $\bar c$ and $c$ are Grassmann variables whose Fourier transformation we define as $\bar c_{\nu,\sigma} = \frac{1}{\sqrt{\beta}}\int_0^\beta d\tau\; \bar c_{\tau,\sigma}\;e^{-i\nu\tau}$ and $c_{\nu,\sigma} = \frac{1}{\sqrt{\beta}}\int_0^\beta d\tau\; c_{\tau,\sigma}\;e^{i\nu\tau}$ where $\nu=(2n+1)\pi/\beta,n\in\mathbb{Z}$ are fermionic Matsubara frequencies. 
Further, the static (bosonic transfer frequency $\omega=0$) generalized susceptibility in charge channel is defined as
\begin{align}
\begin{split}
\label{eq:defChi}
    \chi_\text{c}^{\nu\nu^\prime}[G_0^{-1}] &= \beta \fdv{G_\nu[G_0^{-1}]}{G_{0,\nu^\prime}^{-1}}\\
    &=\frac{\beta}{Z}\int \mathcal{D}[\Bar{c},c]\; \frac{1}{2} \sum_\sigma\bar c_{\nu,\sigma} c_{\nu,\sigma} \sum_{\sigma^\prime}\bar c_{\nu^\prime,\sigma^\prime} c_{\nu^\prime,\sigma^\prime}\; e^{-S}\\
    &\quad-2\beta G_\nu G_{\nu'}
    .
\end{split}
\end{align}

\textsl{Appendix B: Generalization to non-local, SU(2)-broken, multi-orbital systems --}
In the main text, the formalism is worked out for local, SU(2)-symmetric, single-orbital systems. Here, we want to give a sketch on how to generalize the method to non-local, SU(2)-broken, multi-orbital systems. First, we define the iterative map for this case as
\begin{align}
\begin{split}
    f_{G_0,k,\sigma}^{\alpha\beta}\left[G_0^{-1}\right] = &p\,([G_{k,\sigma,\text{phys}}^{-1}]^{\alpha\beta}+\Sigma_{k,\sigma}^{\alpha\beta}[G_0])\\
    &+(1-p)\,\left[G_{0,k,\sigma}^{-1}\right]^{\alpha\beta},
\end{split}
\end{align}
with $k=(\nu,k_x,k_y,k_z)$, spin $\sigma$ and $\alpha,\beta$ being orbital indices (we still assume energy, momentum and spin conservation).
The Jacobian can be written as
\begin{align}
    J_{G_0}^{ij} = \fdv{f_{G_0,i}}{G_{0,j}^{-1}} = \delta_{ij}-p[\chi_0^{-1}]_{il}\chi^{lj},
\end{align}
where we group $(k,\sigma,\alpha,\beta)$ and $(k^\prime,\sigma^\prime,\alpha^\prime,\beta^\prime)$ in super-indices $i$ and $j$. 
Then $J_{G_0}$ can be viewed as a matrix in this superindex space and $G_0^{-1}$ and $f_{G_0}$ are vectors in this superindex space. The matrices for the generalized susceptibilities $\chi_{ij}$ and $[\chi_0^{-1}]_{ij}$ are defined as 
\begin{align}
\begin{split}
    \left[\chi_0^{-1}\right]_{ij} &= \left[\chi_0^{-1}\right]_{k\sigma\alpha\beta,k^\prime\sigma^\prime\alpha^\prime\beta^\prime} = -\frac{1}{\beta} \left[G^{-1}_{k\sigma}\right]^{\alpha\alpha^\prime} \left[G^{-1}_{k\sigma}\right]^{\beta^\prime\beta}\delta_{kk^\prime}\delta_{\sigma\sigma^\prime},\\
    \chi^{ij} &= \chi_{k\sigma\alpha\beta,k^\prime\sigma^\prime\alpha^\prime\beta^\prime} = \beta \frac{\delta G^{\alpha\beta}_{k\sigma}}{ \delta[G^{-1}_{0,k^\prime\sigma^\prime}]^{\alpha^\prime\beta^\prime}}.
\end{split}
\end{align}
Therefore, all following steps can be proceeded analogously to the main text by just extending the fermionic Matsubara frequency to the super-index space.
In the following, we want to investigate the spin degree of freedom in more detail. For this, we start by introducing two bases, the spin-index-basis [left-hand side of \cref{eq:basis}] and the charge-spin-basis [right-hand side of \cref{eq:basis}], which are transformed into each other by $S$ \cite{bickers2004a}
\begin{align}
\label{eq:basis}
\begin{pmatrix}
\chi_{\uparrow\uparrow} & \chi_{\uparrow\downarrow}\\
\chi_{\downarrow\uparrow} & \chi_{\downarrow\downarrow}
\end{pmatrix}
=S\begin{pmatrix}
\chi_\text{c} & \chi_\text{cs}\\
\chi_\text{sc} & \chi_\text{s} 
\end{pmatrix}S \quad\text{with } S=\frac{1}{\sqrt{2}}\begin{pmatrix} 1&1\\1&-1\end{pmatrix},
\end{align}
where $\chi_{\sigma\sigma^\prime}=\fdv{G_\sigma}{G_{0,\sigma^\prime}^{-1}}$, $\chi_\text{c/s} = \left(\chi_{\uparrow\uparrow}+\chi_{\downarrow\downarrow}\pm(\chi_{\uparrow\downarrow}+\chi_{\downarrow\uparrow})\right)/2$ and $\chi_\text{cs/sc}=\left(\chi_{\uparrow\uparrow}-\chi_{\downarrow\downarrow}\pm(-\chi_{\uparrow\downarrow}+\chi_{\downarrow\uparrow})\right)/2$. 
In order to use the charge-spin-basis in our context, we also need to transform the iterative map into this basis
\begin{align}
    \begin{pmatrix}
        f_\uparrow\\f_\downarrow 
    \end{pmatrix} 
    \xlongrightarrow{S} 
    \frac{1}{\sqrt{2}}
    \begin{pmatrix}
        f_\uparrow+f_\downarrow\\f_\uparrow-f_\downarrow
    \end{pmatrix}
\end{align}
Enforcing SU(2)-symmetry, as it is done in the main text, implies $f_\uparrow=f_\downarrow=:f$. Then expanding the iterative map in the charge-spin-basis around the fixed point with a small perturbation $\eta$ gives
\begin{align}
\begin{split}
\label{eq:fix_p_cs_pert}
    \begin{pmatrix}
        \eta^{(n+1)}\\0
    \end{pmatrix}=&\left[\mathbb{1}-p
    \begin{pmatrix}
        \chi_0^{-1}\chi_\text{c} & 0\\
        0 & \chi_0^{-1}\chi_\text{s} 
    \end{pmatrix}\right]
    \begin{pmatrix}
        \eta^{(n)}\\0
    \end{pmatrix}\\
    &+\mathcal{O}\left((\eta^{(n)})^2\right).
\end{split}
\end{align}
From \cref{eq:fix_p_cs_pert}, it becomes clear why only the charge channel contributes to the Jacobian in the main text. Note, however, that if one does not suppress the spin degree of freedom, the statistical noise from, e.g., Monte Carlo can be enough to trigger the instabilities in the spin channel also for SU(2)-symmetric parameters.

\textsl{Appendix C: Functional representation of the Jacobian}
We want to emphasize that the Jacobians of the iterations in $G_0$, [\cref{eq:J_g0}], and $G$, [\cref{eq:J_g}], are very different regarding their functional form. To make this more visible, we explicitly write the functional dependence in the following. By using \mbox{$\Sigma\left[G_0^{-1}\right] = G^{-1}_{0,\nu}-G^{-1}_\nu[G^{-1}_0]$} and \cref{eq:defChi}, we get
\begin{align}
    J_{G_0}^{\nu\nu^\prime}\left[G_0^{-1}\right] =\fdv{f_{G_0,\nu}}{G_{0,\nu^\prime}^{-1}} = \delta_{\nu\nu^\prime}-p\frac{\chi_\text{c}^{\nu\nu^\prime}\left[G_0^{-1}\right]}{-\beta G^2_\nu\left[G_0^{-1}\right]}
\end{align}
and with the help of $\delta\Sigma_\nu/\delta G_{\nu'}=\beta\Gamma^{\nu\nu'}_\text{c}$ we obtain
\begin{align}
\begin{split}
    J_G^{\nu\nu^\prime}\left[G^{-1}\right] = \fdv{f_{G,\nu}}{G_{\nu^\prime}^{-1}} =(1-p)\delta_{\nu\nu^\prime}+p\,\Gamma_\text{c}^{\nu\nu^\prime}\left[G\right]\, \beta G^2_{\nu^\prime}
\end{split}
\end{align}
where the Bethe-Salpeter equation $\Gamma_\text{c} = \chi_\text{c}^{-1}-\chi_0^{-1}$
was used for \cref{eq:J_g} in the main text. Hence, even though the functional form of both Jacobians is different, both expressions 
can be brought to a very similar form.

\textsl{Appendix D: Practical implementation and numerical efficiency of the modified iteration --}
In our procedure, the Jacobian to stabilize the physical solution should be evaluated at the (in principle unknown) physical fixed point.  For the data presented in this work (e.g.~\cref{fig:stay_g0}) the analytically known matrices for $\chi_0^{-1}\chi_\text{c}$, computed exactly at the considered parameter set, have been used to construct $\mathcal{P}$. 
However, we want to emphasize that, in practice, an {\sl a priori} knowledge of the physical fixed point is \emph{not} needed in order to apply our modified scheme to more realistic, non analytically solvable cases. 
The crucial insight is that all quantities needed for defining our modified scheme evolve smoothly by tuning $U$ (with the obvious exception of phase transitions).
Therefore,  $\chi_0^{-1}\chi_\text{c}$ and/or $\mathcal{P}$ do not need to be evaluated exactly at each specific value of $U$ at which the calculation is performed (see SM~\cite{SM} for a detailed discussion). This can also be used to reduce the overall numerical cost considerably by evaluating  $\chi_0^{-1}\chi_\text{c}$  on a coarser $U-$grid than the one-particle quantity. 
We also note that a further reduction of numerical cost for our modified scheme can be obtained (i) by computing $\chi_0^{-1}\chi_\text{c}$ on a smaller Matsubara frequency grid than the one-particle quantity, and (ii) by calculating only few eigenvalues instead of the full spectra of $\chi_0^{-1}\chi_\text{c}$ e.g.~with Arnoldi iteration \cite{arnoldi1951,lehoucq1998}, provided that the unstable eigenvalues are included (see SM~\cite{SM} for more detail).
Our finding that several misleading convergence issues, previously attributed to the branching of the LWF, can be in fact be ascribed to intrinsic stability conditions of the physical fixed point of the iterative schemes considered pave the route to systematically explore the performances of other iterative algorithms able to converge to the physical fixed point. One concrete possibility, beyond the procedure presented here, is offered by (quasi-)Newton methods, for which we provide a first preliminary comparison w.r.t. our scheme. Specifically, a more in depth comparison is done for a specific quasi-Newton method, namely the Chord method, and our modified iteration in the SM~\cite{SM}.

\textsl{Appendix E: Flow-diagram for iterations --}
The Jacobian only gives information about the stability of the fixed point in its respective local neighborhood. 
In the worst case, this could mean that the fixed point is only stable for starting points in an infinitesimal neighborhood around the fixed point, which would be practically useless. Therefore it is also important to gain information about the non-local (in the space of the iterated quantity, e.g., $G_0^{-1}$) behavior around the fixed point. For this, we will define two quantities: first,
\begin{align}
    \Delta G_{0,\nu}^{-1}[G_{0}^{-1}] = f_{G_{0},\nu}[G_{0}^{-1}] - G_{0,\nu}^{-1},
\end{align}
which represents the change between two iterations and second the Lyapunov function $V$ \cite{guckenheimer1983,strogatz2015,argyris2017,chen2023}, which gives information about the stability. 
A Lyapunov function needs to fulfill the following conditions: (i) $V[\bar G_0^{-1}]=0$ where $\bar G_0^{-1}$ is a fixed point of $f_{G_0}$, (ii) $V[G_0^{-1}]> 0$ for $G_0^{-1}\neq\bar G_0^{-1}\in\mathcal{D}$ where $\mathcal{D}$ is an open domain around $\bar G_0^{-1}$, (iii) $V$ is continuous in $\mathcal{D}$, (iv) $V[G_0^{-1}]\rightarrow \infty$ as $\norm{G_0^{-1}}\rightarrow \infty$.
$\bar G_0^{-1}$ is globally (over the entire domain $\mathcal{D})$ and asymptotically stable in $\mathcal{D}$ if $\Delta V\left[[G_0^{(n)}]^{-1}\right]:=V\left[[G_0^{(n+1)}]^{-1}\right]-V\left[[G_0^{(n)}]^{-1}\right]<0$ for all $[G_0^{(n)}]^{-1}\neq\bar G_0^{-1}$ in $\mathcal{D}$.
Therefore $\mathcal{D}$ refers to a \emph{stability region} that consists of initial points which converge to the fixed point $\bar G_0^{-1}$ \cite{chen2023}. 
For our purpose, we define $V$ as
\begin{align}
    V[G_{0}^{-1}] := \norm{f_{G_{0}}[G_0^{-1}]-G_{0}^{-1}}^2,
\end{align}
where $\norm{.}$ denotes the usual Euclidean norm for complex vector spaces. Since the frequencies in the ZP model are decoupled, one can view all quantities essentially as complex scalars. Therefore we adapted $V_\nu[G_{0}^{-1}] := |f_{\nu,G_{0}}[G_0^{-1}]-G_{\nu,0}^{-1}|^2$ when considering the $\nu^\text{th}$ frequency.
A visualization of this tool is done in \cref{fig:stay_g0} (middle panel) in the main text for the ZP model. There $\Delta G_{0,\nu}^{-1}$ is plotted as a flow-diagram in the complex plane of $G_{0,\nu}^{-1}$. The color grading of this plot is done with $\Delta V$ where red means $\Delta V[G_0^{-1}]>0$ and blue means $\Delta V[G_0^{-1}]<0$. We can therefore consider the blue region around a stable fixed point as the \emph{stability region} $\mathcal{D}$ around a stable fixed point.
Note that this concept is worked out explicitly for the iterative scheme in $G_0$ but can be also done for the iteration in $G$ by simply changing $G_0\rightarrow G$ and $f_{G_0}\rightarrow f_G$.
\end{document}

\title{Origin of misleading convergence in self-consistent many electron theories:\\Fundamental aspects and practical implications\\
\emph{--Supplemental Material--}}

\author{Herbert E{\ss}l\orcidlink{0009-0005-9883-8104}}\email{herbert.essl@tuwien.ac.at}
\TUVienna
\author{Matthias Reitner\orcidlink{0000-0002-2529-0847}}
\TUVienna
\author{Evgeny Kozik\orcidlink{0000-0001-6580-9570}}
\Kings
\author{Alessandro Toschi\orcidlink{0000-0001-5669-3377}}
\TUVienna

\date{\today}

\maketitle
Within this Supplemental Material, we provide more details referring to the analytically known quantities of the zero point (ZP) model in \cref{sec:models}.
In \cref{sec:BM_ZP_comp} we elaborate on the physical relevance of the ZP model and in \cref{sec:it_map_ZP} we give additional information on the iterative maps in the ZP model.
In \cref{sec:G0,sec:bold} we show additional data for the conventional and modified iteration for $G_0$ and $G$, for both the HA and the ZP model.
In \cref{sec:practicaluse} we offer guidance on how to implement the modified scheme in praxis.
In \cref{sec:comp_newton} we compare our introduced modified iteration to (quasi-)newton methods, where a more in depth comparison is done for the Chord method. 
%Finally , in \cref{sec:data} we give a link to a repository where the data and scripts that are used for this work are available.

\section{\label{sec:models}Additional information for the zero point model}
Here, we explicitly provide the functional forms of $G_0,\,G,\,\Sigma$ and $\chi_\text{c}$ for the ZP model which are also given in Ref.~\cite{rossi2015}
\begin{align}
\label{eq:G0_ZP}
    &G_{0,\text{phys},\nu} = \frac{1}{i\nu+\delta\mu},\\
\label{eq:G_ZP}
    &G_\nu[G_0^{-1}] = \frac{G_{0,\nu}^{-1}}{G_{0,\nu}^{-2}-U/\beta},\\
\label{eq:SigmaG0_ZP}
    &\Sigma_\nu[G_0] = \frac{U}{\beta}G_{0,\nu},\\
\label{eq:SigmaG_ZP}
    &\Sigma_{\text{w/s},\nu}[G] = \frac{2UG_\nu/\beta}{1\pm\sqrt{1+4UG_\nu^2/\beta}},\\
\label{eq:chi_ZP}
    &\chi^{\nu\nu^\prime}_\text{c}[G_0^{-1}] =\beta\fdv{G_\nu}{G_{0,\nu^\prime}^{-1}}= -\beta\frac{G_{0,\nu}^{-2}+U/\beta}{(G_{0,\nu}^{-2}-U/\beta)^2}\delta_{\nu\nu^\prime}.
\end{align}
Note that we have employed a slightly modified version of the ZP, differing from the one originally presented in Ref.~\cite{rossi2015}. Following Ref.~\cite{kim2020a}, which connects the $N-$replica ZP model to the HA, we introduced (decoupled) fermionic Matsubara frequencies for the ZP model to make the comparison between ZP model and HA more transparent. 
However, since the introduced frequencies are decoupled we can still regard the one- and two-particle Green's functions as complex scalars as it is for the original ZP model.

The values of $U$ where the branch cut appears, i.e., $\Sigma_\text w[G_\text{phys}]=\Sigma_\text{phys}$ for $U<U_\text{branch}$ and $\Sigma_\text s[G_\text{phys}]=\Sigma_\text{phys}$ for $U>U_\text{branch}$, and the $U-$values of the \mbox{(pseudo-)}divergences for $\chi_0^{-1}\chi_\text{c}$ are given by
\begin{align}
    U_\text{branch}&=\pm\abs{G^{-1}_{0,\nu}}^2,\\
    U_\text{div}&=-\beta\frac{(\delta\mu^2+\nu^2)^2}{\delta\mu^2-\nu^2},
\end{align}
respectively. Here, we find that a vertex divergence only appears at $\delta\mu=0$, and otherwise, we find pseudo-divergences.

The respective expressions for the HA are also known analytically (but not in a functional form) they can be found in Refs.~\cite{pairault2000,essl2024}.

\section{Physical relevance of the zero point model}\label{sec:BM_ZP_comp}
\begin{figure}[t!]
\includegraphics[scale=0.4]{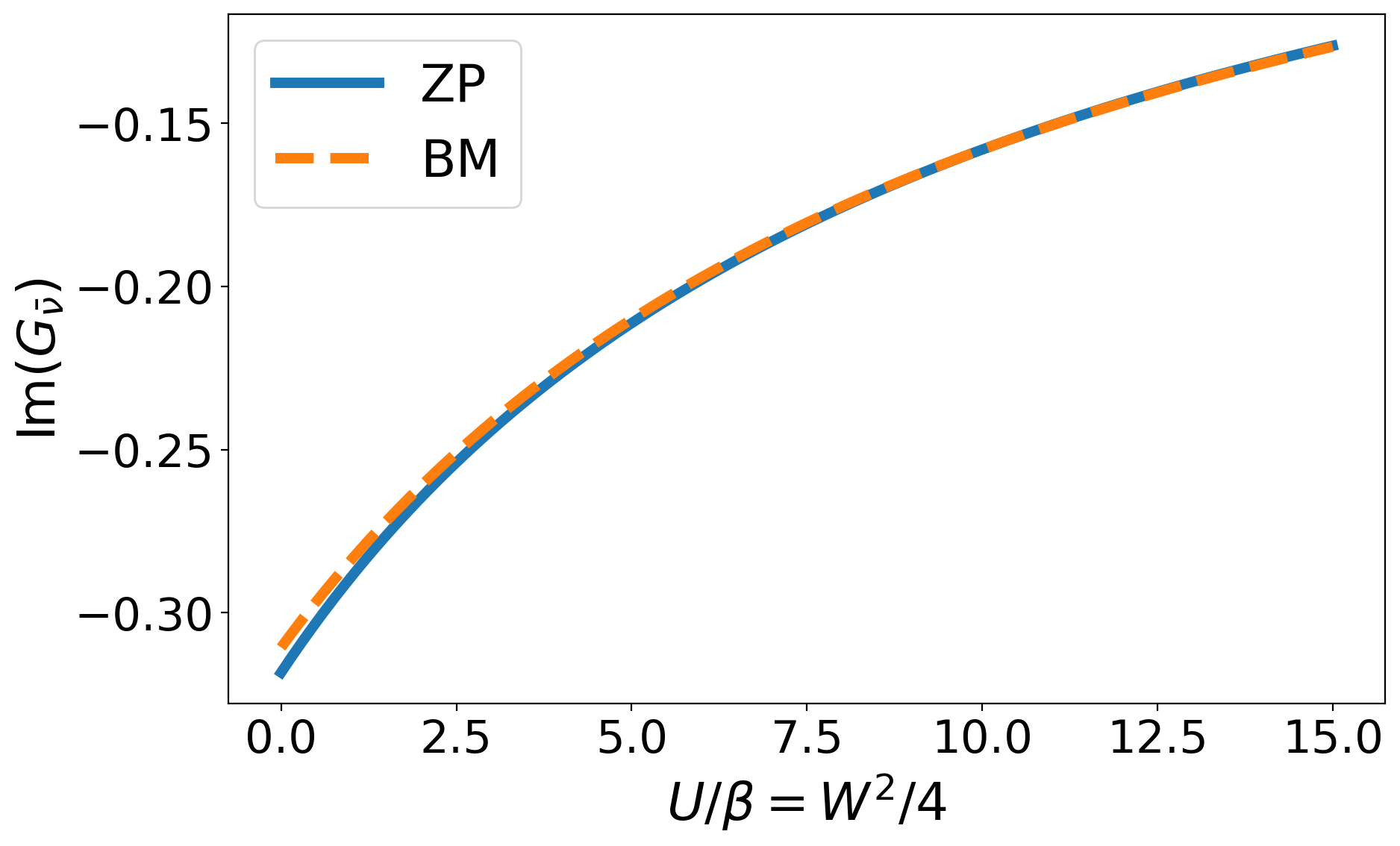}
\caption{\label{fig:BM_ZP_compare}Imaginary part of the physical Green's functions $G_{\bar\nu}$ at $\bar\nu=\pi/\beta$ at ph-symmetry for the BM and the ZP model. It can be seen that when equaling $U/\beta$ of the ZP model with $W^2/4$ of the BM the two Green's functions are almost equivalent.} 
\end{figure}

While the ZP \emph{per se} does not describe a specific realistic system, a physical interpretation of its results, as well as of its non-perturbative features such as vertex divergences, can nonetheless be given, because of its direct connection to the binary mixture (BM) problem. In the latter, one considers a disordered lattice system, where each site is randomly associated to one of of two possible energy values (whose difference is $W$). The direct link with the BM problem is provided by the fact that the coherent potential approximation (CPA) of the BM in the limit of of high lattice coordination number (i.e.~where DMFT becomes exact) and particle-hole symmetry yields essentially the same LWF functional of the ZP model. This can be seen by looking at the functional form of the self-energy which was calculated in Ref.~\cite{schafer2016}: $\Sigma_{\text{w/s}\nu}^\text{BM}[G]=\frac{W^2 G_\nu/2}{1\pm\sqrt{1+W^2 G_\nu^2}}$. Evidently, the functional form of $\Sigma[G]$ in the ZP model is equivalent by equaling the prefactor ($U/\beta$) of the four point term of the ZP action with $\frac 14 W^2$. Further we find that the physical Green's functions of the BM and the ZP model are nearly the same, where the agreement is even better for strong coupling which is the region of interest for this manuscript. The Green's functions for both models at $\bar \nu=\pi/\beta$ are shown in \cref{fig:BM_ZP_compare}.

As a result, the two models are essentially the same, and, thus, the large $U$ regime of the ZP model at ph-symmetry can be indeed associated to the onset of a disorder-driven insulating phase in the BM problem (solved by CPA).
While the BM problem holds greater physical relevance, its parametrization away from particle-hole symmetry is ambiguously. Consequently, from a formal perspective the ZP model is preferred, as it provides both an explicitly defined action functional and a clear parametrization away from particle-hole symmetry.

\section{Additional information for the iterative maps of the ZP model}\label{sec:it_map_ZP}
\begin{figure}[t!]
\includegraphics[scale=0.35]{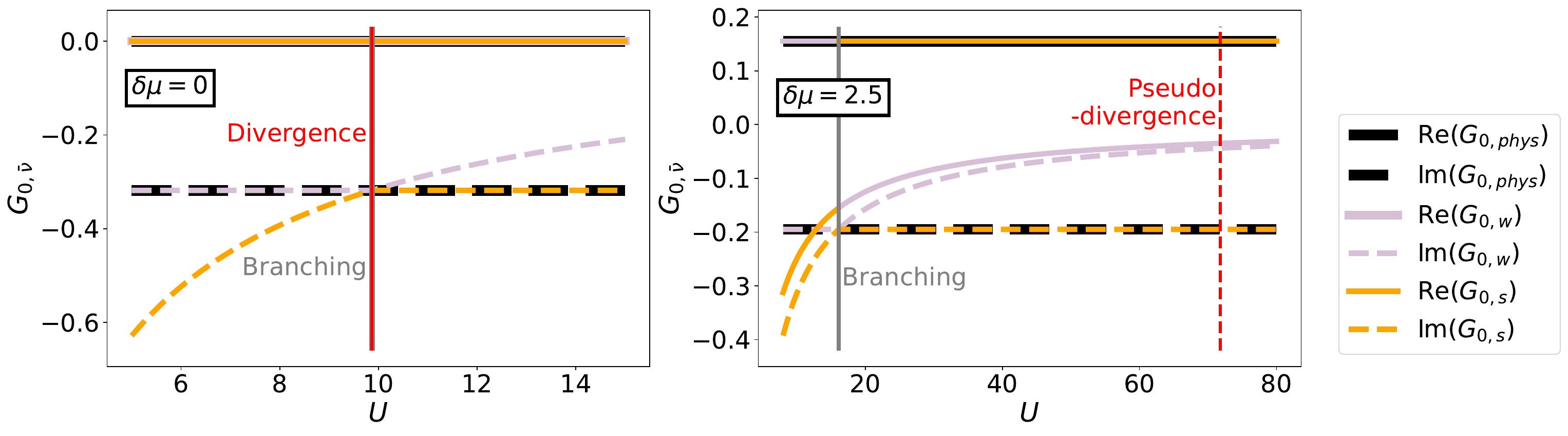}
\caption{\label{fig:fixed_points}The weak and strong coupling fixed points $G_{0,\text{w}}$ and $G_{0,\text{s}}$ of the iterative map $f_{G_0}$ at $\bar\nu=\pi/\beta$ together with the physical value of $G_0$ plotted at ph-symmetry (left panel) and out of ph-symmetry (right panel)} 
\end{figure}

Investigating the iterative map $f_{G_0}$ of the main text we find the fixed points by solving $f_{G_0}[G_0^{-1}]=G_0^{-1}$ analytically for the ZP model. We find that $f_{G_0}$ for the ZP model has the two fixed points
\begin{align}
    G_{0,\text{w/s},\nu}^{-1}[G] = G^{-1}_\nu + \Sigma_{\text{w/s},\nu}[G]=\frac{2G_\nu U/\beta}{-1\pm\sqrt{4G_\nu^2U/\beta+1}},
\end{align}
which are plotted for $G_\nu=G_{\text{phys},\bar\nu}=\frac{i\bar\nu+\delta\mu}{(i\bar\nu+\delta\mu)^2-U/\beta}$ with $\bar\nu=\pi/\beta$ in \cref{fig:fixed_points}.
Note that the branching and the divergence lie on top of each other (left panel), while the pseudo-divergences line of $\chi_0^{-1}\chi_\text{c}$ appears after the branch point at $U_\text{div}>U_{\text{branch}}$ (right panel).
Further, the branching between $G_{0,\text{w}}$ and $G_{0,\text{s}}$ is only continuous in the case of a
divergence while it is continuous in the imaginary part and discontinuous in the real part if a pseudo-divergence is present.
For the ZP model, also the $U$-value where the physical fixed point gets unstable can be analytically calculated.
For the iteration in $G_0^{-1}$, we solve the equation $\abs{1-p\chi_0^{-1}\chi_\text{c}}^2=1$, which gives two solutions:
one positive 
$U_{\text{break},G_0,-}>0$, and one negative $U_{\text{break},G_0,+}<0$:
\begin{align}
\begin{split}
\label{eq:Ubreak_g0}
    &U_{\text{break},G_0,\pm} = \beta\frac{p-2}{\frac{2 \nu ^2 (p-1)}{\left(\delta\mu ^2+\nu ^2\right)^2}+\frac{1-p}{\delta\mu ^2+\nu ^2}\pm\frac{\sqrt{\frac{p^2 \left(\delta\mu ^4-2 \delta\mu ^2 \nu ^2 (2 (p-2) p+1)+\nu ^4\right)}{\left(\delta\mu ^2+\nu ^2\right)^4}}}{p}}.
\end{split}
\end{align}
For $U>0$, which is considered in our studies of the main text, 
$U_{\text{break},G_0,-}$ for $p=1$ is equal to $U_\text{branch}$ (we suspect that this is a peculiarity of the ZP model). 
Since the absolute value of $U_{\text{break},G_0}$ increases monotonically when decreasing $p$, we conclude that the instability can also occur at higher absolute $U$-values then the branching. 
This insight shows again that the branching and the misleading convergence are two different aspects of the problem, which we could disentangle more formally.

For the iteration in $G$, we need to solve the equation $\abs{1-p\chi_0\chi^{-1}_\text{c}}^2=1$, which gives 
\begin{align}
\label{eq:Ubreak_g}
    U_{\text{break},G} = \beta\frac{(p-2) \left(\delta\mu ^2+\nu ^2\right)^2}{2 \left(\delta\mu ^2-\nu ^2\right)}.
\end{align}
Furthermore, there is a finite region around the \mbox{(pseudo-)}divergence for which the physical fixed point is unstable also for the modified iterative scheme of the main text (if $p>0$). For the iteration in $G$, this region exists for both pseudo-divergences and divergences, while it exists only for pseudo-divergences for the iteration in $G_0$.
The lower bound of this region $U_\text{mod}^\text{low}$ is equal to $U_\text{break}$ of \cref{eq:Ubreak_g0,eq:Ubreak_g} since the modified scheme is equivalent to the conventional one before the (pseudo-)divergence. The upper bound $U_\text{mod}^\text{up}$ can be calculated by solving the equation $\abs{1-\mathcal{D}^{\alpha\alpha}\lambda_\alpha}^2=1$ for the iteration in $G_0$ and $\abs{1-\mathcal{D}^{\alpha\alpha}/\lambda_\alpha}^2=1$ for the iteration in $G$. 
In the ZP model this amounts to solving the equation \mbox{$\abs{1+p\chi_0^{-1}\chi_\text{c}}^2=1$} for the iteration in $G_0$ resp.~$\abs{1+p\chi_0\chi^{-1}_\text{c}}^2=1$ for the iteration in $G$. Therefore we just have to replace $p$ with $-p$ in \cref{eq:Ubreak_g0,eq:Ubreak_g} to get 
\begin{align}
\begin{split}
\label{eq:upper_mod_g0}
    &U_{\text{mod},G_0,\pm}^\text{up} = \beta\frac{-p-2}{\frac{2 \nu ^2 (-p-1)}{\left(\delta\mu ^2+\nu ^2\right)^2}+\frac{1+p}{\delta\mu ^2+\nu ^2}\pm\frac{\sqrt{\frac{p^2 \left(\delta\mu ^4-2 \delta\mu ^2 \nu ^2 (2 (-p-2) (-p)+1)+\nu ^4\right)}{\left(\delta\mu ^2+\nu ^2\right)^4}}}{-p}}
    \end{split},\\
    &U_{\text{mod},G}^\text{up} = -\beta\frac{(p+2) \left(\delta\mu ^2+\nu ^2\right)^2}{2 \left(\delta\mu ^2-\nu ^2\right)}.
\end{align}
For $p\rightarrow0^+$ and $U>0$, both $U_{\text{break},G_0,-}$ and $U^\text{up}_{\text{mod},G_0,+}$ approach the divergence at $U_\text{div}$ ($U_{\text{break},G_0,-}$ from below and $U^\text{up}_{\text{mod},G_0,+}$ from above), while for $U<0$, $U_{\text{break},G_0,+}$ and $U^\text{up}_{\text{mod},G_0,-}$ approach $U_\text{div}$.
Evidently, also $U_{\text{break},G}$ and $U^\text{up}_{\text{mod},G}$ approach $U_\text{div}$ from below and above (for $p\rightarrow 0^+$).
This shows that the region, where the physical fixed point is unstable for the modified scheme can be made arbitrarily small with $p$.

\section{Additional data on $G_0$ iteration}\label{sec:G0}
\begin{figure}[t!]
\includegraphics[scale=0.6]{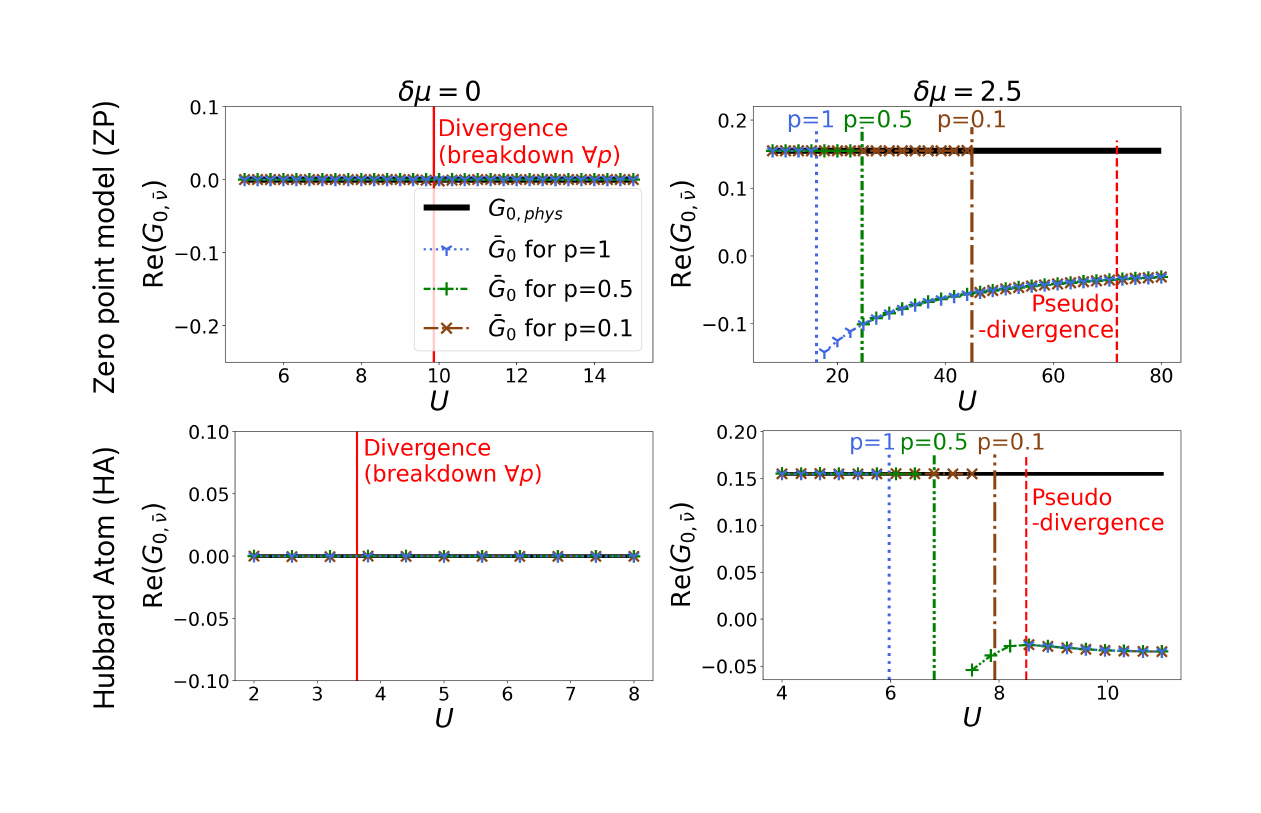}
\caption{\label{fig:mis_conv_re}Misleading convergence conventional $G_0$ iteration in the ZP model (upper panels) and HA (lower panels).
The real part of $\bar G_{0,\bar\nu=\pi/\beta}$ obtained via the conventional iteration is shown and compared to the physical solution $G_{0,\text{phys},\bar\nu}$ at ph-symmetry (left panels) and out of ph-symmetry (right panels).} 
\end{figure}

In this section we provide additional data to the onset of the misleading convergence for $f_{G_0}$ for the same parameters as shown in the main text in Fig.~2. In \cref{fig:mis_conv_re}, we show the real part of $G_0$ at $\bar\nu=\pi/\beta$.
We find that, at ph-symmetry, the real part of $G_0$ is zero also for the unphysical solution for the ZP model and the HA (left panels of \cref{fig:mis_conv_re}). 
Out of ph-symmetry for a parameter where a pseudo-divergence appears, we find that the real part of $G_0$ switches discontinuously between the physical and the unphysical fixed point (right panels of \cref{fig:mis_conv_re}), this discontinuity is necessary according to the proof given in Ref.~\cite{gunnarsson2017} which states that a crossing of a physical and an unphysical solution implies a vertex divergence. 
The imaginary part of $G_0$ is continuous at some point between the instability at $p=1$ and the pseudo-divergence for both models (see Fig.~2 in the main text). 

\begin{figure}[t!]
\includegraphics[scale=1]{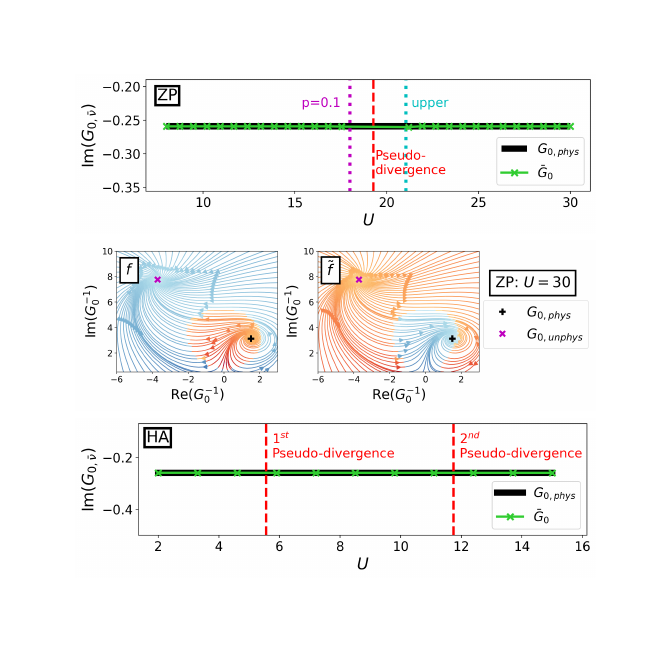}
\vspace{-12mm}
\caption{\label{fig:stay_SM}Im($G_{0,\bar\nu}$) obtained with the modified iterative scheme $\tilde f$ out of ph-symmetry ($\delta\mu=1.5, p=0.1$) for both the ZP model (top panel) and the HA (bottom panel). The comparison with the physical solution demonstrates the stabilization of the physical fixed point in the whole parameter regime. Middle panel: flow-diagram in the complex plane of $G_0^{-1}$ for the ZP model in the nonperturbative regime ($U =30$) for the conventional (left) and the modified scheme (right).} 
\end{figure}

Further, we show that the modified scheme $\tilde f_{G_0}$ also works for a type (iii) instability which is connected to a pseudo-divergence of $\chi_0^{-1}\chi_\text{c}$. 
In \cref{fig:stay_SM} the same type of calculations as in Fig.~3 in the main text are shown, but for $\delta\mu=1.5$ where a pseudo-divergence is present. 
In addition, also $U_{\text{break},G_0,-}$ and $U^\text{up}_{\text{mod},G_0,+}$ are marked for the ZP model by a magenta and cyan line to explain the region around the pseudo-divergence where $\tilde f_{G_0}$ does not converge to the physical fixed point. As explained before, this region would vanish for $p\rightarrow 0^+$.

\section{Additional data on $G$ iteration\label{sec:bold}}
\begin{figure}[t!]
\includegraphics[scale=0.35]{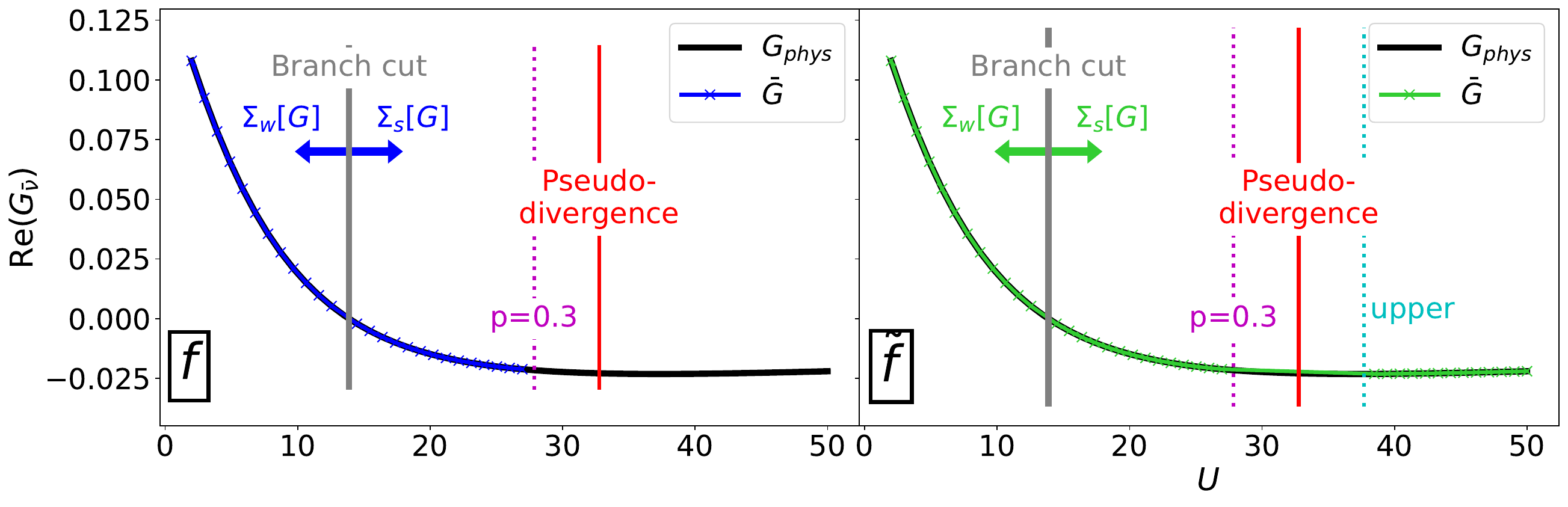}
\caption{\label{fig:bold_re}Real part of $G_{\bar\nu}$ for the iteration in $G$ for the ZP model out of ph-symmetry ($\delta\mu=2, p=0.3$) using the conventional scheme (left) and using the modified scheme (right). $U_{\text{break},G}$ and $U^\text{up}_{\text{mod},G}$ are marked by a magenta and cyan line.}
\end{figure}

In \cref{fig:bold_re}, the real parts of $G$ for calculations of Fig.~4 in the main text ($\delta\mu=2$) are shown.
Additionally, to $U_{\text{break},G}$ (magenta line) we now also mark $U^\text{up}_{\text{mod},G}$ by a cyan line to explain the region of non convergence for $\tilde f_G$ in the lower panel. This region would again vanish for $p\rightarrow 0^+$.

\begin{figure}[t!]
\includegraphics[scale=0.35]{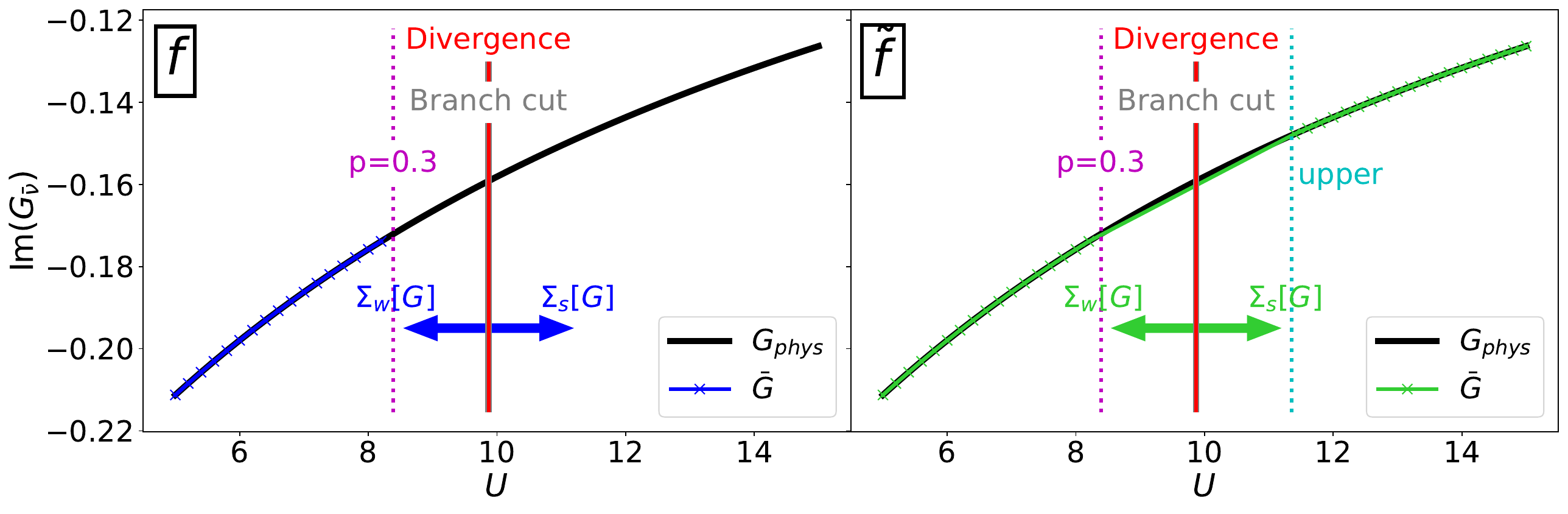}
\caption{\label{fig:bold_dmu0}Similar to \cref{fig:bold_re}, now for the imaginary part of $G_{\bar\nu}$ for the iteration in $G$ for the ZP model at ph-symmetry ($\delta\mu=0, p=0.3$).}
\end{figure}

\begin{figure}[t!]
\includegraphics[scale=0.35]{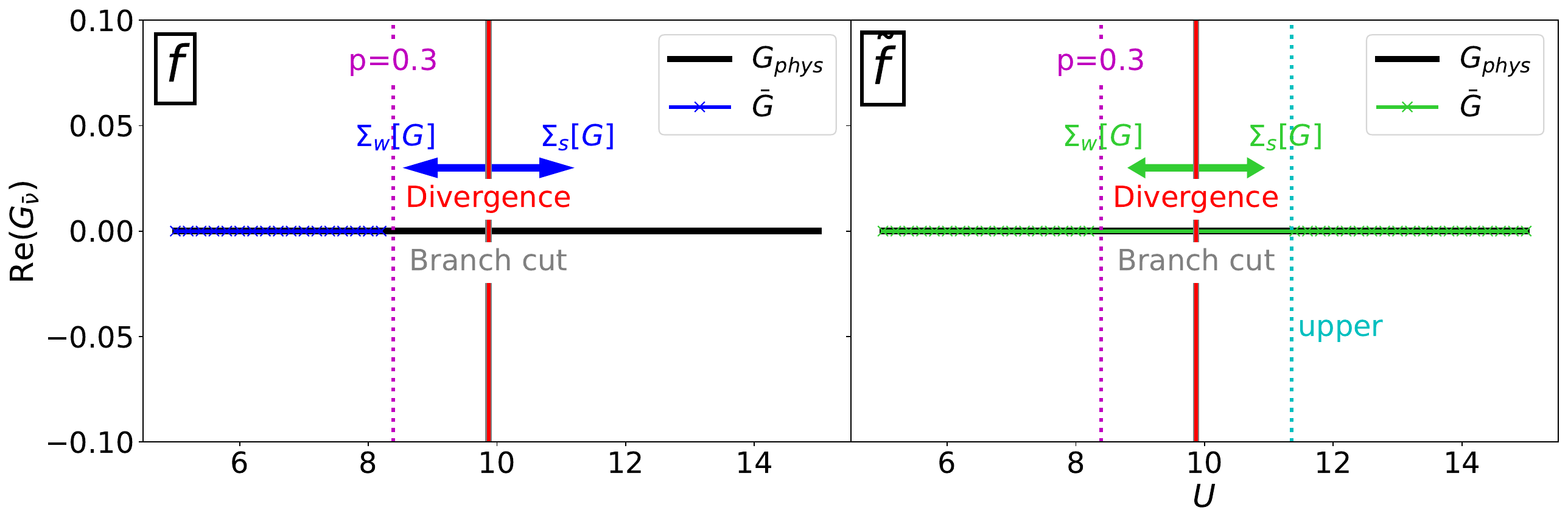}
\caption{\label{fig:bold_re_dmu0}Same as \cref{fig:bold_dmu0}, but for the real part of $G_{\bar\nu}$.}
\end{figure}

Further, we also show data for $\delta\mu=0$ where a vertex divergence is present, to demonstrate that $\tilde f_G$ also works for this case. In \cref{fig:bold_dmu0} the imaginary part of $G$ is shown while we show its real part in \cref{fig:bold_re_dmu0}.

\section{Guidance on practical/efficient implementation of the modified scheme \label{sec:practicaluse}}
As discussed in appendix D of the main text, the results presented in this work (\cref{fig:stay_SM} and Fig.~3 in the main text) have been obtained exploiting the analytically known matrices for $\chi_0^{-1}\chi_\text{c}$, computed exactly at each considered parameter set, to construct $\mathcal{P}$. 
In a more realistic situation, this would require the knowledge of the (in general unknown) physical fixed point to evaluate the corresponding Jacobian. However, as concisely discussed in appendix D, since all quantities needed for defining the modified scheme evolve continuously as a function of the model parameter (e.g. $U$), in practice, this problem can be naturally circumvented.

For instance, we outline in the following a 
possible implementation of the modified scheme for cases, in which the solution of $\chi_0^{-1}\chi_\text{c}$ is not known {\sl a priori}.
One should start from a parameter region, where the 
physical fixed point is known to be  stable (e.g. at a small enough $U$-values). 
By gradually approaching the parameter region of interest (e.g., by gradually increasing $U$) one can easily compute the Jacobian at the fixed-point after calculating it by the iterative procedure. Doing this, e.g. for selected points in the parameter space, requires to perform a single two-particle calculation \emph{only} at the very end
of the iterative procedure. This will allow to quickly track whether any eigenvalue of the Jacobian is approaching an absolute value of 1 in the parameter region considered. 
As the absolute values of the eigenvalues of the Jacobian will never exceed 1 in the conventional/non-modified scheme (as the unphysical fixed-point are stable after the crossing), the evolution of the problematic Jacobian eigenvalue will display a characteristic cusp-like dependence as a function of $U$ (or a more general parameter) around the point, where its absolute value reaches $1$.
This thus allows to readily identify the parameter values, where the modified scheme might become necessary and how this should be defined.  Indeed, if the misleading-convergence problems cannot be simply fixed by more damping, the modified scheme is easily implemented by changing $D^{\alpha\alpha}$ in $\mathcal{P}$ from $p$ to $-p$ for the problematic eigenvalue. Then the physical fixed point will be automatically stabilized in the adjacent larger $U$ region.
This specific calculation protocol has been successfully tested by using the eigenvectors of $\chi_0^{-1}\chi_\text{c}$ at fixed $U=2$ to construct $\mathcal{P}$ for the entire displayed $U$-range ($U=2-15$). 

It is also important to mention that the numerical cost for the practical implementation of the modified scheme sketched above can be, in general, further reduced by exploiting proper approximations for the Jacobian calculation.

First, since the eigenvectors associated to the instabilities are in general localized at low Matsubara frequencies \cite{thunstrom2018a,essl2024} one can calculate the Jacobian (or the corresponding two-particle quantities) on a much smaller frequency grid than the one used for the Green's function and construct the associated $\mathcal{P_\text{small}}$ on that low-frequency grid. 
Then $\mathcal{P}$ on the large frequency is readily computed as
\begin{align}
    \mathcal{P} = \begin{pmatrix}
                p\mathbb{1}&\mathbb{0}&\mathbb{0}\\
                \mathbb{0}&\mathcal{P_\text{small}}&\mathbb{0}\\
                \mathbb{0}&\mathbb{0}&p\mathbb{1}
                \end{pmatrix}.
\end{align}
This approximation has been tested for one of the cases discussed in the main text ($\delta\mu=0, U=2-15,\beta=1$, Fig.~3 of the main text). Specifically, in this test calculation the Green's function was computed on $N=150$ positive Matsubara frequencies, while the Jacobian calculation was restricted to 30 positive frequencies. The converged results obtained in this way are, nonetheless, identical to the ones presented in the main text.

Another approximation strategy, which we also tested, is to exploit the Arnoldi method (an extension of the Lanczos algorithm to general, possible non-Hermitian, matrices) \cite{arnoldi1951,lehoucq1998} instead of computing the full eigenspectrum of the Jacobian. 
In the realm of the Arnoldi method, one can compute only $k<2N$ eigenvalues and eigenvectors of the Jacobian. This yields a $2N\times k$ matrix $\mathcal{U}$ of eigenvectors. In this context, one can thus define the $\mathcal{D}$ matrix [cf. in Eq.~(5) of the main text] as a $k\times k$ matrix, where the condition if the diagonal entry of $\mathcal{D}$ is $p$ or $-p$ is given by the sign of the real part of the eigenvalues obtained from the Arnoldi algorithm. Then, the modified iterative scheme can be readily defined through $\mathcal{P}=\mathcal{UDU}^+$ where $\mathcal{U}^+$ denotes the Moore-Penrose inverse \cite{campbell1991} of $\mathcal{U}$.  Evidently, by doing this, one also needs to re-include the information on the (typically conventional) high-frequency eigenvectors of the Jacobian, mostly neglected in the Arnoldi space.
This can be achieved, e.g., by noting that
 $\mathcal{U^+U}=\mathbb{1}$ but $\mathcal{UU^+}\neq\mathbb{1}$ since $\mathcal{U}$ has linear independent columns. 
Evidently, the fact that $\mathcal{UU^+}\neq\mathbb{1}$ is problematic since,  if no unstable eigenvalue is present (as it typically happens at low $U$ values),  $\mathcal{D}=p \mathbb{1}$ and therefore also $\mathcal{P}=p\mathcal{UU}^+=p \mathbb{1}$ should hold.
Hence, in order to accurately reconstruct the high-frequency/conventional part of the Jacobian, one can apply the following procedure to $\mathcal{P}$: Identify the indices $i$ where $\mathcal{UU^+}$ is not close to 1 on the diagonal and then simply set for these coefficients $\mathcal{P}_{ii}$ equal to $p$.
Clearly, in order to use the Arnoldi method to speed-up the implementation of the modified scheme, it is essential that all eigenvalues with negative real part are captured in the Arnoldi method. 
Provided that \mbox{$k \ge \#$ of eigenvalues with negative real part}, this poses no problems, since exactly these eigenvalues are captured first by Lanczos-like methods. 
Indeed, by applying the Arnoldi method to the same testbed case mentioned above (this time $G_0$ and $\chi_0^{-1}\chi_c$ are both calculated for $N=150$ positive fermionic Matsubara frequencies), where we had at most 4 negative eigenvalues of $\chi_0^{-1}\chi_c$, we have used $k=6$ for a calculation and found that this approximation reproduces the results where the exact $\chi_0^{-1}\chi_c$ is used.

\section{Comparison of the modified scheme to (quasi-)Newton methods\label{sec:comp_newton}}
As demonstrated in the main text, the misleading convergence problems do not necessarily originate only from the multivaluedness of the LWF, but also from the instability of the physical fixed point. While no general solution is known for the former problem, in the latter case, several algorithmic strategies can be adopted. 
One of the most direct approach to this problem is the proposed modified algorithm introduced in our work.
This does not represent, however, the only possible strategy. Another reasonable choice in the presence of multiple fixed points would be, for instance, to resort to (quasi)-Newton methods, as these iterative schemes can, in principle, access (depending on the initial conditions) all fixed points.

While it is not possible to make a definite statement about which of these two procedures is going to perform better (since this will depend on several details in the implementation, as well as on the specific iterative algorithm, which needs to be stabilized), we intend to discuss  here some general advantage/disadvantages of the both approaches in this specific context.

We start by noticing that an evident benefit of the modified iterated scheme presented in the manuscript is that the Jacobian needed for the stabilization needs to be calculated only \emph{once} for each point in the phase space. This is different w.r.t.~the Newton method, where the Jacobian (or an approximation thereof, like in quasi-Newton methods) must be calculated at each iteration step.
We note that this aspect is particular important, since the iterative map is usually not known analytically and therefore the calculation of the Jacobian may become numerically costly.

A further, useful feature of the proposed method is that it can be used to filter out the unstable/unphysical fixed points in a natural, controlled manner.
This might be an advantage over the Newton method, which finds all roots of $f(x)-x$,  and, hence, both stable and unstable fixed points of $f(x)$ ($f$ corresponds to the iterative map [Eqs.~1 and 2] of the main text and $x$ to the argument of this map, i.e.~$G_0^{-1}$ or $G^{-1}$). In that framework, the starting point of the iterations is going to determine to which of the multiple fixed points of $f(x)$ the Newton method will eventually converge. 
The damped iteration scheme, per construction, can only converge, instead, to stable fixed points of $f(x)$ (if it converges). This property can be used to construct an iterative scheme where one can converge \emph{only} to the physical fixed point essentially in the whole parameter regime.
To this aim, we recall that for the conventional damped iteration the physical fixed point is stable at small interaction, becoming typically unstable for increasing interactions, at a bifurcation point.
If we first consider type (i) instabilities (the eigenvalue that causes the instability is real), we observe that, at the bifurcation point the physical and unphysical fixed point exchange their stability nature, i.e. the physical fixed point becomes unstable  with increasing interaction while the unphysical one becomes stable. 
In this case, our modified damped iteration scheme, which flips the stability criterion at the bifurcation, can only converge, as desired, to the physical fixed point in the whole parameter regime\footnote{This argumentation can be formally supported, by looking at a bifurcation analysis. We suspect that the type (i) instability is a \emph{transcritical bifurcation}, at which it is known that the two fixed points that cross at the bifurcation exchange stability, see e.g. \cite{strogatz2015}. 
In this context, if the iteration scheme is modified in such a way that the matrix element $\mathcal{D}^{\alpha\alpha}$ of the eigenvalue, that will become unstable, is always set to $-p$ (and not only if $\text{Re}(\lambda_\alpha)<0$), the stability of the two fixed points will be reversed, i.e.~the physical fixed point becomes stable, while the unphysical one becomes unstable by increasing interaction. Then, if one always chooses the iteration scheme for which the physical fixed point is stable one constructs, de facto, our proposed modified iteration approach.}.
The case of type (iii) instabilities (the eigenvalue that triggers the instability is complex) appears formally more involved\footnote{Indeed, we suspect that these instabilities correspond to Neimark-Sacker bifurcations. Here, the physical fixed point becomes unstable and a stable (quasi) orbit arises going from small to large $U$ values.}, but, eventually, does not pose additional difficulties in practice. In that case, we observe that there is some small region where both the physical and the unphysical solution are stable, which in the ZP model happens in the proximity of the breakdown of the skeleton series.
However, we find (empirically) that, with the modified scheme, only the physical branch is stable at sufficiently large $U$ values. This leads again to a controlled way of convergence, through the stabilization of the physical fixed point, in essentially the whole phasespace.
Finally, it is not problematic that there exists a region where both the physical and the unphysical solutions are stable, as the corresponding fixed points do not intersect because there is no vertex divergence \cite{gunnarsson2017}.

At the same time, a clear systematic advantage of the Newton method is the faster convergence (quadratic order of convergence vs.~ the linear order of convergence of the damped scheme). Another one is that (quasi-)Newton methods have been intensively studied in the past: They are thus well documented and allow for several professional optimizations.
Hence,  it might be turn very useful, in the future, depending on the specific stabilization algorithm considered, to combine a (quasi-)Newton procedure together with numerical continuation \cite{allgower1990,veltz:hal-02902346} (which can be used to follow the most regular path at bifurcation points). This would represent a promising, complementary path to stabilize the physical fixed point in the whole parameter regime in the most efficient way.

As a first step in this direction we tested a specific quasi-Newton method named Chord method for the $G_0^{-1}$ iteration in the HA. In the Chord method the Jacobian is fixed during the iteration, usually the Jacobian is evaluated at the starting point of the iteration. 
To get a first comparison between the Chord method and our modified scheme we use the former converged result as a starting point and the Jacobian evaluated at the physical solution at the former set of parameters (for the modified scheme we precede as described in \cref{sec:practicaluse}). 

\begin{figure}[t!]
    \centering
    \includegraphics[width=\linewidth]{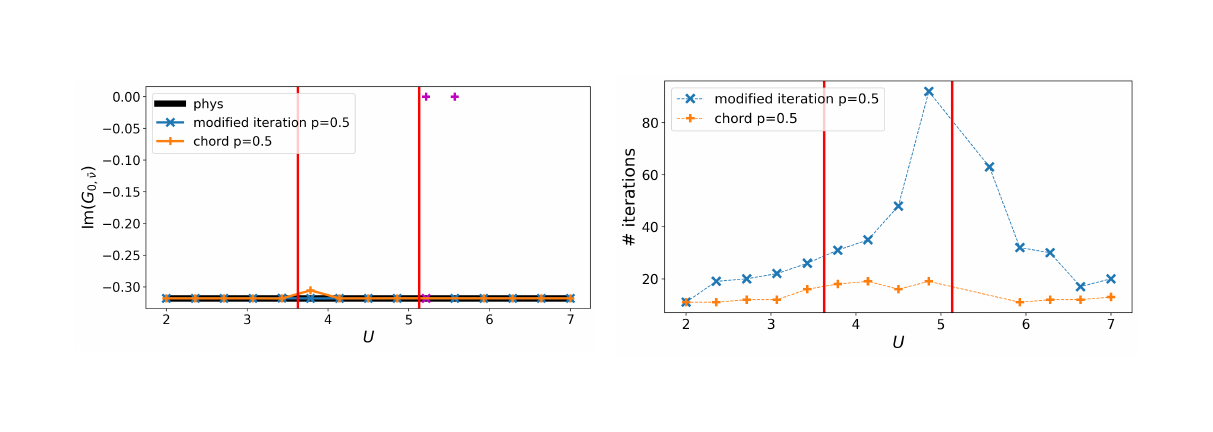}
    \vspace{-1.5cm}
    \caption{Chord and modified iteration results for  $G_0^{-1}$ of the HA at ph-symmetry (with damping $p=0.5$). The previous converged solution is taken as starting point. Left panel: $\text{Im}(G_{0,\bar\nu})$ of the converged solution for both methods in comparison to the physical one. Magenta markers show not converged results. Right panel: Number of iterations required for convergence.}
    \label{fig:lr_chord_vs_modified}
\end{figure}

In the parameter range considered, we find that the chord method, when it converges, appears indeed to converge faster than the modified iteration scheme (right panel of \cref{fig:lr_chord_vs_modified}).
At the same time, as we show in the left panel of \cref{fig:lr_chord_vs_modified}, the chord method converges to the unphysical fixed point for the parameter data set right after the first vertex divergence. Arguably, this problem is likely not just restricted to the parameter region in the very immediate proximity after the vertex divergence, if one would compute the Jacobian by performing an actual 2-particle calculation, as it is required in more realistic situations, where its exact physical expression is not known {\sl a priori}. We also observe that the chord method displays divergence ($G_0^{-1}\rightarrow\infty$) for the first two parameter datasets after the second vertex divergence, while the modified iteration lies on top of the physical fixed point.

Hence, the chord method, does indeed allow a faster convergence in this case, but, especially in the close proximity of the critical parameter regime, where  vertex divergencies occur, displays more difficulties to converge to the physical fixed point.
\begin{figure}[t!]
    \centering
    \includegraphics[width=\linewidth]{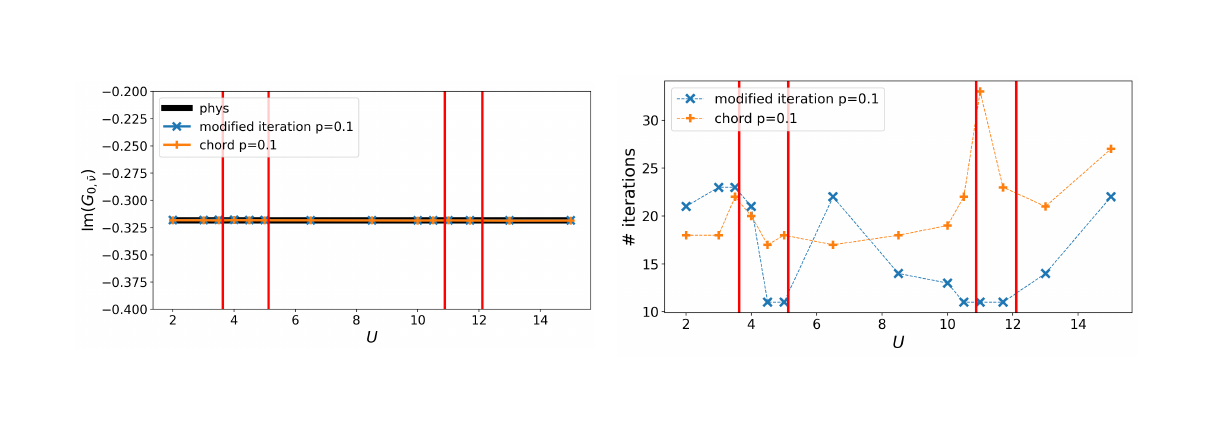}
    \vspace{-1.5cm}
    \caption{Chord and modified iteration scheme, where $G_0^{-1}$ is iterated for the HA at ph-symmetry (with $p=0.1$). The physical solution is taken as starting point, since $G_0^{-1}$ is linear in $U$ this corresponds to a secant predictor (see text). Left panel: $\text{Im}(G_{0, \bar\nu})$ of the converged solution for both methods in comparison to the physical one. Right panel: Number of iterations for the parameter-set, where the corresponding algorithm converged.}
    \label{fig:sec_chord_vs_modified_mix0p1}
\end{figure}
\begin{figure}[t!]
    \centering
    \includegraphics[width=0.5\linewidth]{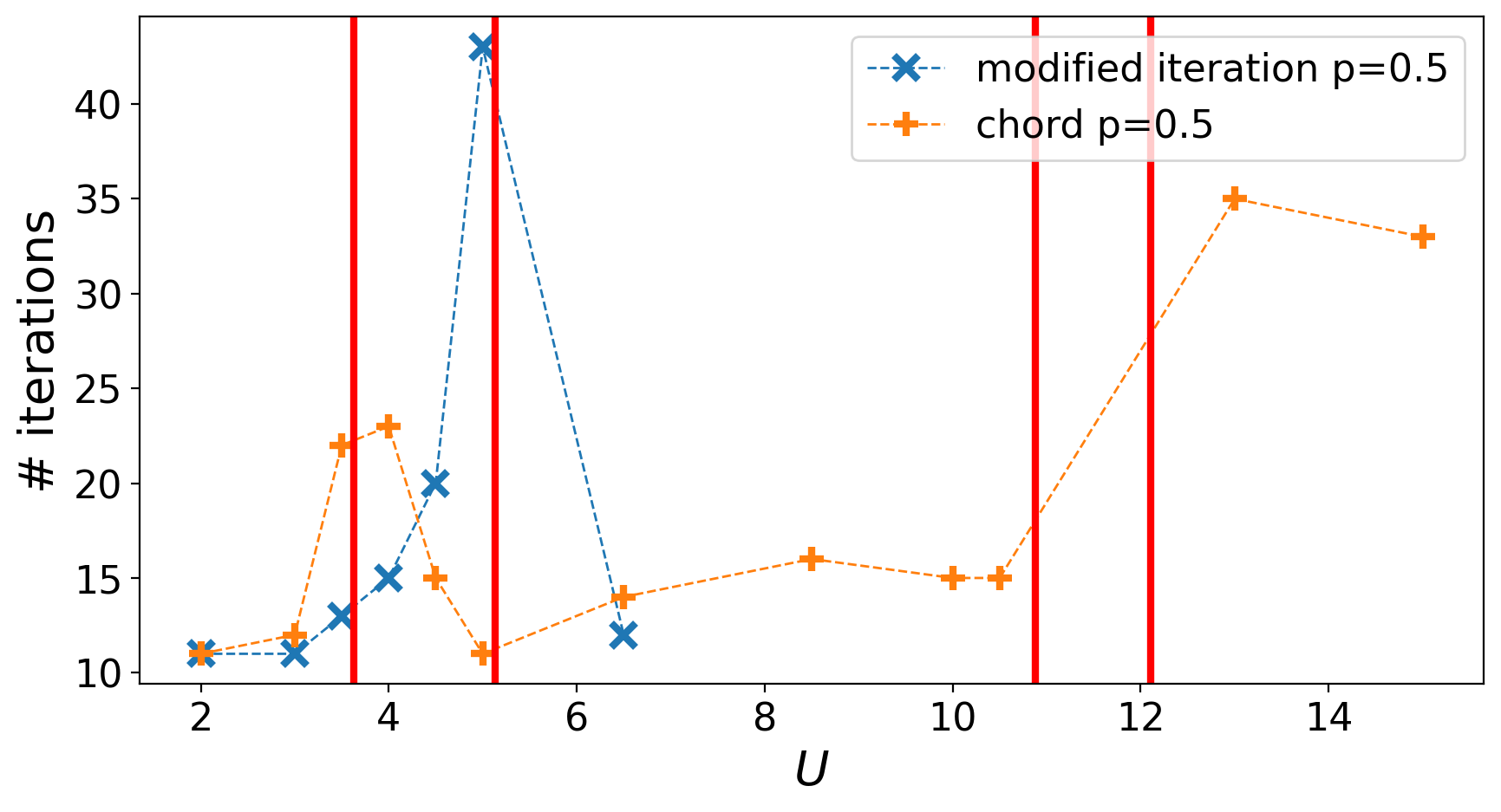}
    \caption{Number of iterations for the same situation as in \cref{fig:sec_chord_vs_modified_mix0p1} but for $p=0.5$}
    \label{fig:it_chord_vs_modified_mix0p5}
\end{figure}

As a next step we use a secant predictor \cite{allgower1990} to approximate the starting point of $G_0^{-1}$, since $G_0^{-1}=i\nu+\delta\mu=i\nu+\mu-U/2$ is linear in $U$ for the HA the secant predictor gives the exact solution. As this blessed situation will, unfortunately, not hold for more complicated/realistic situations, we simulated those cases by starting from a slightly perturbed exact solution. In \cref{fig:sec_chord_vs_modified_mix0p1} we find that both the secant and the modified iteration scheme do converge to the physical solution for all considered datasets (left panel). In this scenario both the Chord method and the modified iteration always converged to the physical fixed point.

However, differently to the situation described in the point above the modified iteration converges faster than the chord method for almost all considered parameter points (right panel).
All calculations displaying this behavior were performed for $p=0.1$.

At the same time, for the sake of completeness, we have also repeated the same calculations  as in \cref{fig:sec_chord_vs_modified_mix0p1}, but for a significantly lower damping ($p=0.5$). In this case, we found that the chord method converges for all considered points, while the modified iteration scheme only converges up to $U<8.5$. 

In conclusion we want to suggest to continue an investigation in this direction to obtain a stable, efficient and practical algorithm.

%\section{Data availability\label{sec:data}}
%A data set containing all numerical data and plot scripts used to generate the figures for this publication is publicly available at \cite{data}.

\bibliography{refs_SM}